\documentclass[useAMS,usenatbib]{mn2e}
\usepackage{txfonts}
\usepackage{graphicx}
\usepackage{amssymb}
\usepackage[below]{placeins}
\usepackage{txfonts}
\usepackage{natbib}
\bibpunct{(}{)}{;}{a}{}{,}

\def \xmm{{\emph{XMM-Newton}}}

\def \chandra{{\emph{Chandra}}}

\newcommand{\ion}[2]{#1\,{\sc{#2}}}

\bibpunct{(}{)}{;}{a}{}{,}

\title[On the thermodynamic self-similarity of relaxed, nearby, giant ellipticals]
{On the thermodynamic self-similarity of the nearest, most relaxed, giant ellipticals}
\author[N. Werner et al.]{N. Werner$^1$\thanks{E-mail: norbertw@stanford.edu}, S. W. Allen$^1$, A. Simionescu$^1$\thanks{Einstein fellow}\\
$^1$Kavli Institute for Particle Astrophysics and Cosmology, Stanford University, 452 Lomita Mall, Stanford, CA 94305-4085, USA \\
and SLAC National Accelerator Laboratory, 2575 Sand Hill Road, Menlo Park, CA 94025, USA \\ 
}
\begin{document}
\maketitle
\begin{abstract}
We present detailed spatially resolved measurements of the thermodynamic properties of the X-ray emitting gas in the inner regions of the five nearest, X-ray and optically brightest, and most X-ray morphologically relaxed giant elliptical galaxies known. Beyond the innermost region at $r\gtrsim1$~kpc, and out to $r\sim6$~kpc, the density, pressure, entropy, and cooling time distributions for the X-ray emitting gas follow remarkably similar, simple, power-law like distributions. Notably, the entropy profiles follow a form $K\propto r^{\alpha}$, with an index $\alpha=0.92$-1.07. The cumulative hot X-ray emitting gas mass profiles and the gas-mass to stellar-light ratios of all five galaxies are also similar. Overall the observed similarity of the thermodynamic profiles in this radial range argues that, in these systems, relativistic jets heat the gas at a similar rate averaged over time scales longer than the cooling time $t_{\rm cool}\gtrsim10^8$ yr. These jets are powered by accretion from the hot gas, or material entrained within it, onto the central super-massive black hole. This jet heating creates an energy balance where heating and cooling are in equilibrium, keeping the hot galactic atmospheres in a `steady-state'. Within $r\lesssim1$~kpc, this similarity breaks down: the observed entropy profiles show well resolved flattening and the values differ from system to system substantially. The accretion rate onto the black hole and the AGN activity, heating the interstellar medium, must therefore vary significantly on time scales shorter than $t_{\rm cool}=10^7$--$10^8$~yr. 

\end{abstract}


\section{Introduction}

\begin{table*}
\caption{Summary of the \chandra\ observations. Columns list the distances of the galaxies, the angular scales per kpc at the corresponding distances, the B-band optical and bolometric X-ray luminosities \citep{osullivan2001}, radio luminosities at 1.4 GHz \citep{condon1998,condon2002}, the line-of-sight Galactic absorbing hydrogen column densities, $N_{\rm H}$, in their directions \citep{kalberla2005}, observation dates and identifiers, detectors, and exposure times after cleaning.  }
\begin{tabular}{lccccccccccc}
\hline\hline
Galaxy  			& Distance & Scale 	 		      & $L_{\rm B}$	   & $L_{\rm X}$   &   $L_{\rm R}$   & $N_{\mathrm{H}}$	&	Obs. date	   	& 	Obs. ID  	&  Detector	& 	Exposure \\
				&(Mpc)	  &   (arcsec kpc$^{-1}$) & ($10^{10} L_\odot$)& ($10^{41}$ erg s$^{-1}$) &($10^{38}$ erg s$^{-1}$)       &     (cm$^{-2}$)		   &				&			&			& (ks)		\\
\hline 
NGC4472 (M49)	&	16.7$^\star$	& 12.4 	       & 8.7 			   & 2.96 		& 1.20 	&	1.53		&	2000 Jun 12	&	321		& ACIS-S3	&	19.1 \\
				&				& 		       & 5.8 			   & & 		& 		&      2010 Feb 27     &      11274       &   ACIS-S3 	 &     39.7 \\
NGC4649 (M60)	&	16.5$^\star$	& 12.5 	       & 			   & 2.05 		& 0.13 	& 	2.04		 &	2007 Jan 30	&	8182		& ACIS-S3	&	49.5\\
				&				& 		       & 			   & 			& 		& 			 &	2007 Feb 01	&	8507		& ACIS-S3	&	17.5 \\
NGC1399 &	20.9$^\star$	& 9.9		       & 4.4 			   & 5.68 		& 1.52 	& 	1.50		&	2000 Jan 18	&	319		& ACIS-S3  	&	49.9	\\
				&				& 		       & 			   & 			& 		& 			&  	2003 May 26	&	4172		& ACIS-I		&	36.7 \\ 
				&				& 		       & 			   & 			& 		& 			&	2008 Jun 08	&	9530		&  ACIS-S3 	&	59.3 \\
NGC1407			&	28.8$^\dagger$& 7.2	       & 7.4 			   & 1.95 		& 1.23  	& 5.42		& 	2000 Aug 16	&	791		&  ACIS-S3 	&	32.5 \\
NGC4261			&	31.6$^\dagger$& 6.5	       & 5.1			   & 1.63 		& 329.52 	& 1.75		&	2008 Feb 12	&	9569		& ACIS-S3  	&	100.9 \\
\hline
\label{obs}
\end{tabular}\\
$^\dagger$\citet{tonry2001}  \\
$^\star$\citet{blakeslee2009}
\end{table*}

X-ray studies with \chandra\ and \xmm\ have shown that relativistic jets, produced by accreting supermassive black holes (SMBH) at the centres of galaxies, groups and clusters, interact strongly with their environments, driving shocks \citep[e.g.][]{forman2005,forman2007,nulsen2005,simionescu2009b,million2010b} and inflating bubbles of relativistic plasma in the surrounding X-ray emitting gas \citep[e.g.][]{churazov2000,fabian2003,fabian2006,birzan2004,dunn2005,dunn2006,dunn2008,forman2005,forman2007,rafferty2006,mcnamara2007}. These bubbles rise buoyantly and can entrain and uplift large quantities of low entropy gas from the innermost regions of their host galaxies \citep{simionescu2008a,simionescu2009a,werner2010,werner2011}.
All of this activity is believed to take place in a tight feedback loop, where the hot interstellar medium (ISM), which is observable in the X-ray band, cools and accretes onto the central SMBH (referred to as active galactic nucleus or AGN), leading to the formation of jets which heat the surrounding gas and drive the above phenomena \citep[for a review see][]{mcnamara2007,gitti2012}. This heating acts to lower the accretion rate, reducing the feedback, until the accretion eventually builds up again. AGN feedback appears to play an important role in regulating the `cooling cores' of galaxy clusters and the formation and evolution of massive galaxies \citep[e.g.][]{peterson2006,croton2006,delucia2007,sijacki2007}. Although much has been learned about AGN feedback, many questions regarding the remarkable balance between heating and cooling remain.

Nearby giant ellipticals are in many respects low redshift proxies for more distant and more luminous cluster cooling cores, allowing us to study AGN feedback `in closeup'. Another key advantage is that the
binding energy per particle is lower in galaxies: a given amount of AGN heating therefore affects the gas in galaxies more obviously than in clusters. By observing a sample of nearby, X-ray bright galaxies with a high degree of completeness, and with data of
sufficient quality to map the detailed thermodynamic properties, we can hope to gain important insight into the physics of AGN feedback, its duty cycle, and its role in galaxy formation. 

One of the most useful thermodynamic quantities in studying the impact of cooling and feedback in galaxies, groups, and clusters of galaxies is the gas entropy which, by rewriting the adiabatic index in terms of X-ray observables, can be defined as $K=kTn^{-2/3}_{\rm e}$, where $kT$ is the temperature and $n_{\rm e}$ is the electron density of the X-ray emitting gas. Because entropy can only be changed by gains and losses of energy, it traces the thermal history of the gas. 
The radial profiles of entropy in statistical samples of galaxy clusters and groups have been analyzed by \citet{pratt2006a,pratt2010}, \citet{cavagnolo2009} and \citet{sun2009b}. However, a detailed study of the thermodynamic properties of giant ellipticals, reaching down to sub-kpc scales, has not been performed. 

Here we present detailed, spatially resolved measurements of the key thermodynamic properties (density, temperature, pressure, entropy, cooling time, and gas mass) of the X-ray emitting gas in the innermost regions of the five nearest, X-ray and optically brightest, and most morphologically relaxed giant elliptical galaxies known. All of these galaxies have deep {\it Chandra} observations, as well as exquisite optical and multi-wavelength data. 

In Sect.~\ref{sample}, we describe our sample, the reduction of the X-ray data, and the spectral analysis. In Sect.~\ref{results}, we present the measurements of both the 2D distributions and the deprojected radial profiles of the key thermodynamic properties. We show, for the first time, that the azimuthally averaged gas density, pressure, cooling time, and entropy profiles of relaxed giant elliptical galaxies are in the radial range $1\lesssim r \lesssim 6$~kpc remarkably similar. Based on these results, in Sect.~\ref{discussion} we present some new insights into the likely thermal history of the X-ray emitting atmospheres of the galaxies. In Sect.~\ref{conclusions}, we summarize our conclusions.

\section{Sample selection, data reduction, and analysis}
\label{sample}
\subsection{Sample selection and its properties}

\begin{figure*}
\begin{minipage}{0.32\textwidth}
\includegraphics[height=6cm,clip=t,angle=0.,bb=36 109 577 683]{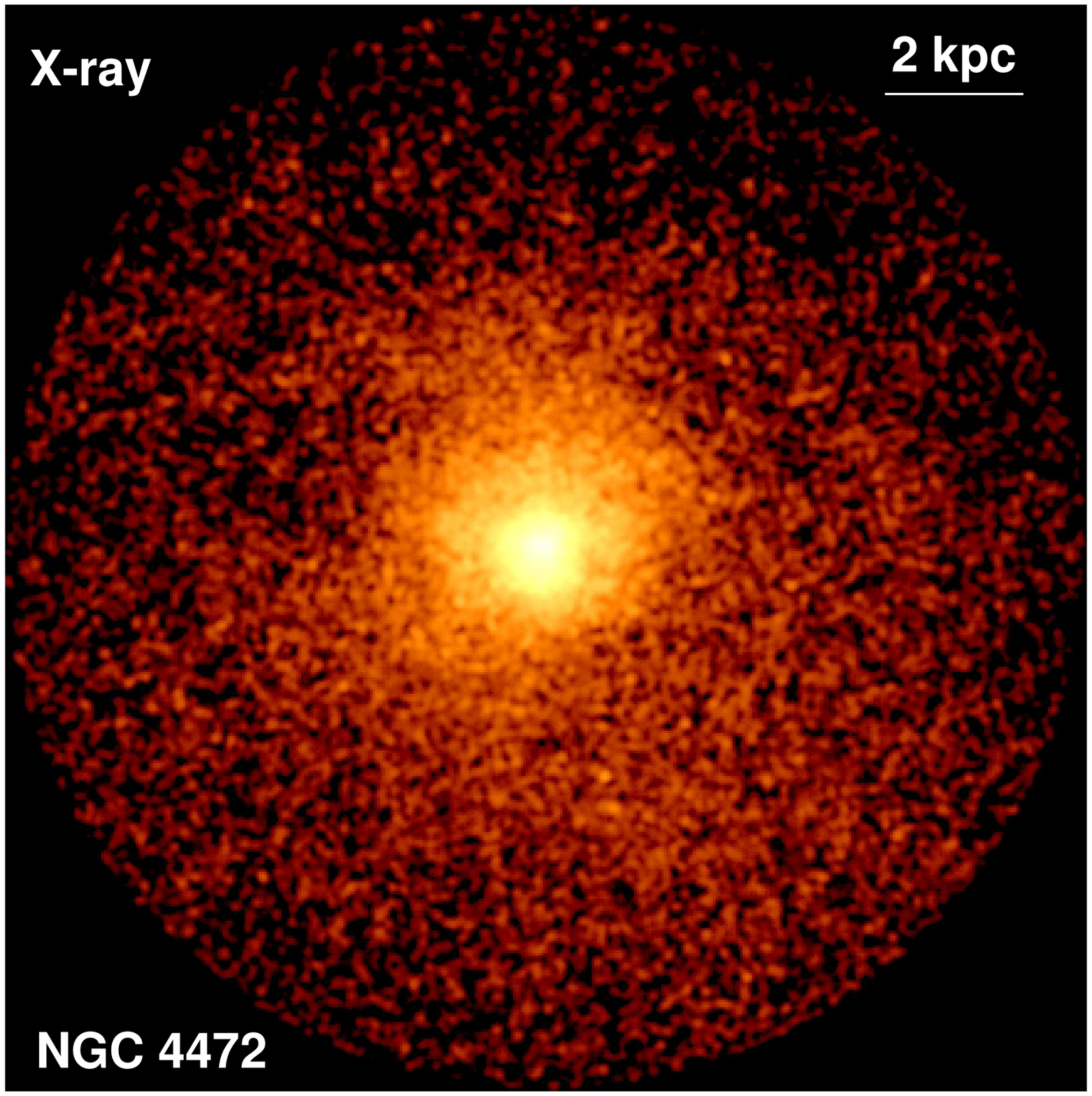}
\end{minipage}
\begin{minipage}{0.32\textwidth}
\includegraphics[height=6cm,clip=t,angle=0.,bb=36 109 577 683]{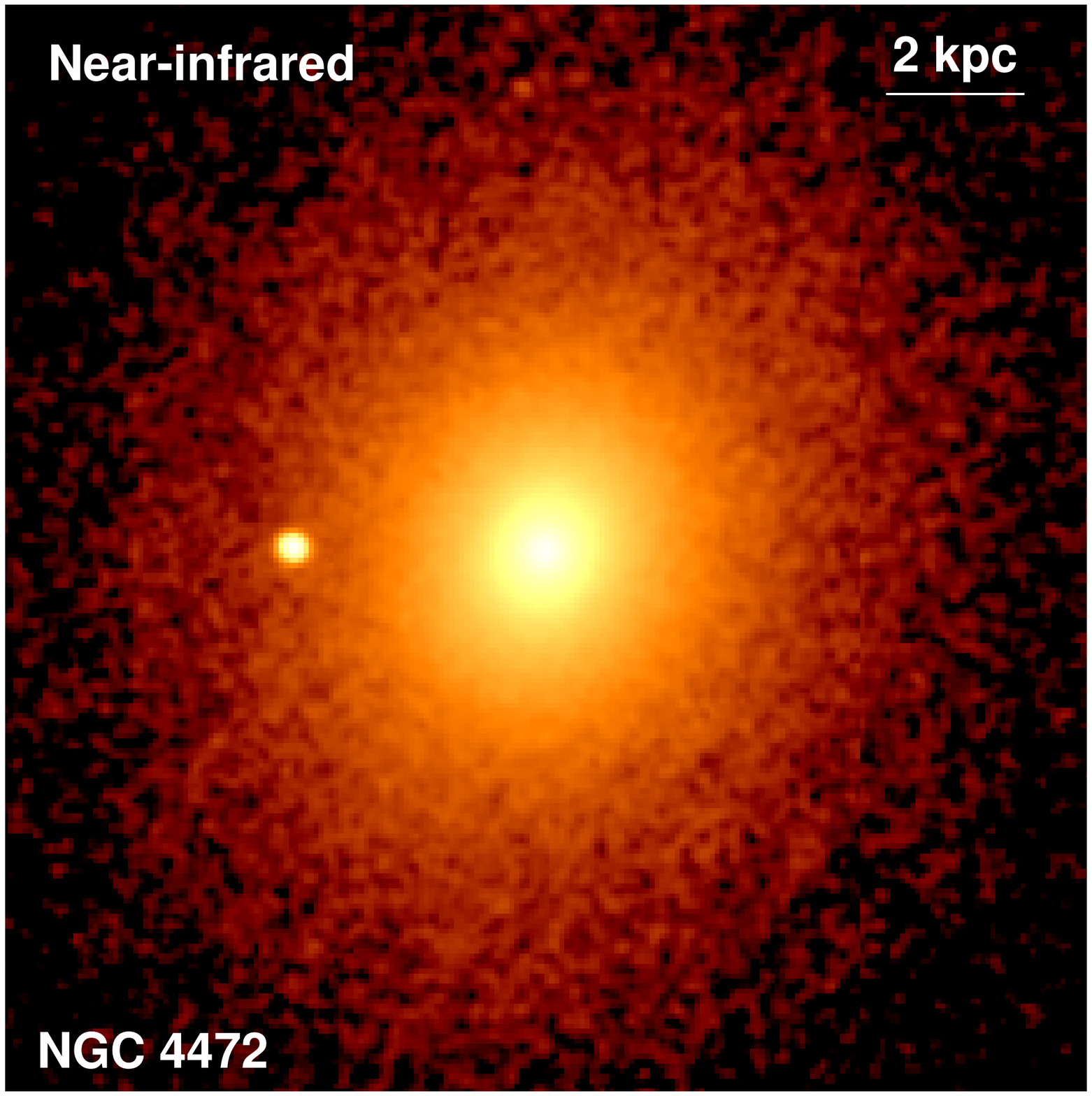}
\end{minipage}
\begin{minipage}{0.32\textwidth}
\includegraphics[height=6cm,clip=t,angle=0.,bb=36 109 577 683]{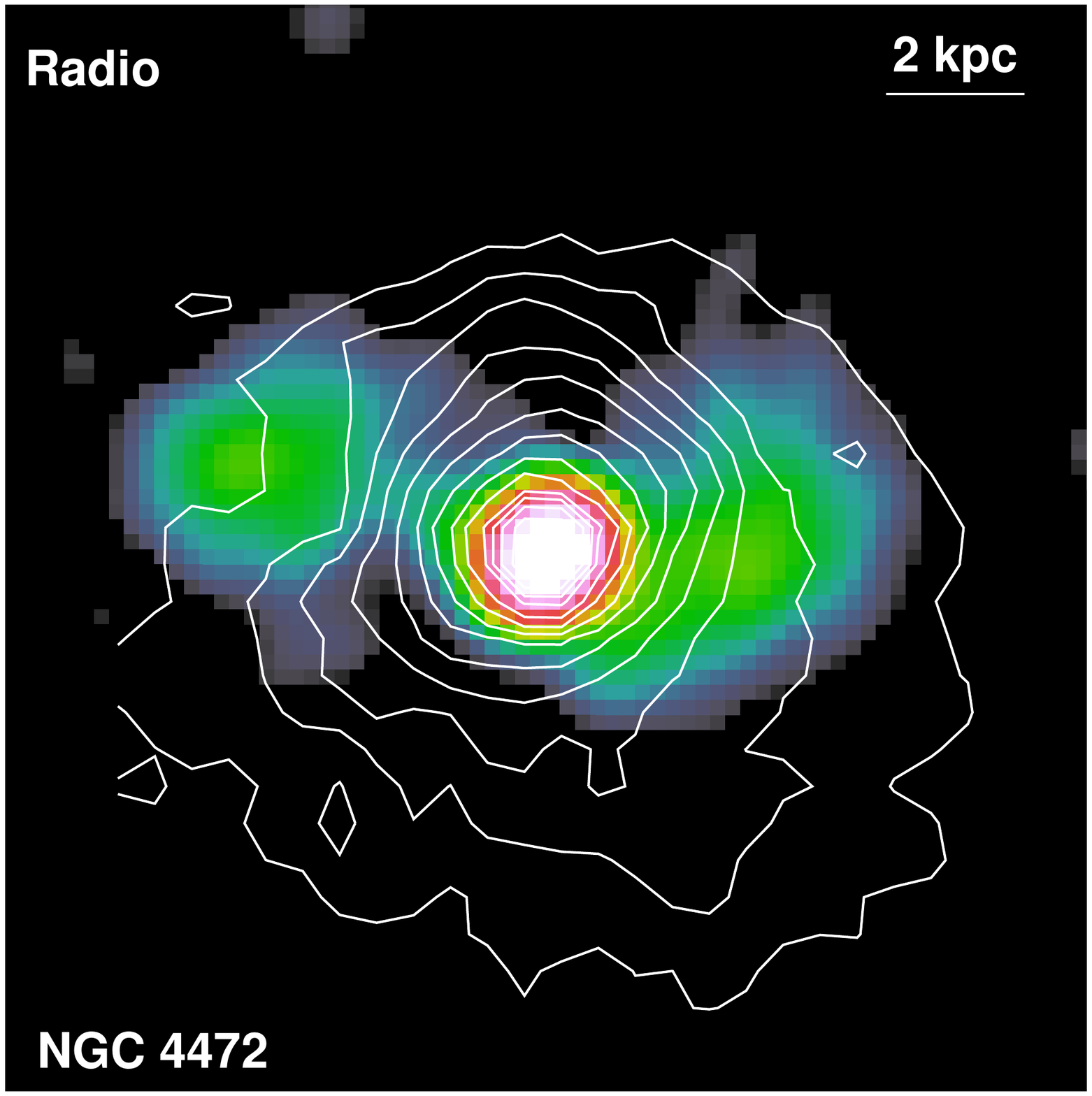}
\end{minipage} \\
\begin{minipage}{0.32\textwidth}
\includegraphics[height=6cm,clip=t,angle=0.,bb=36 109 577 683]{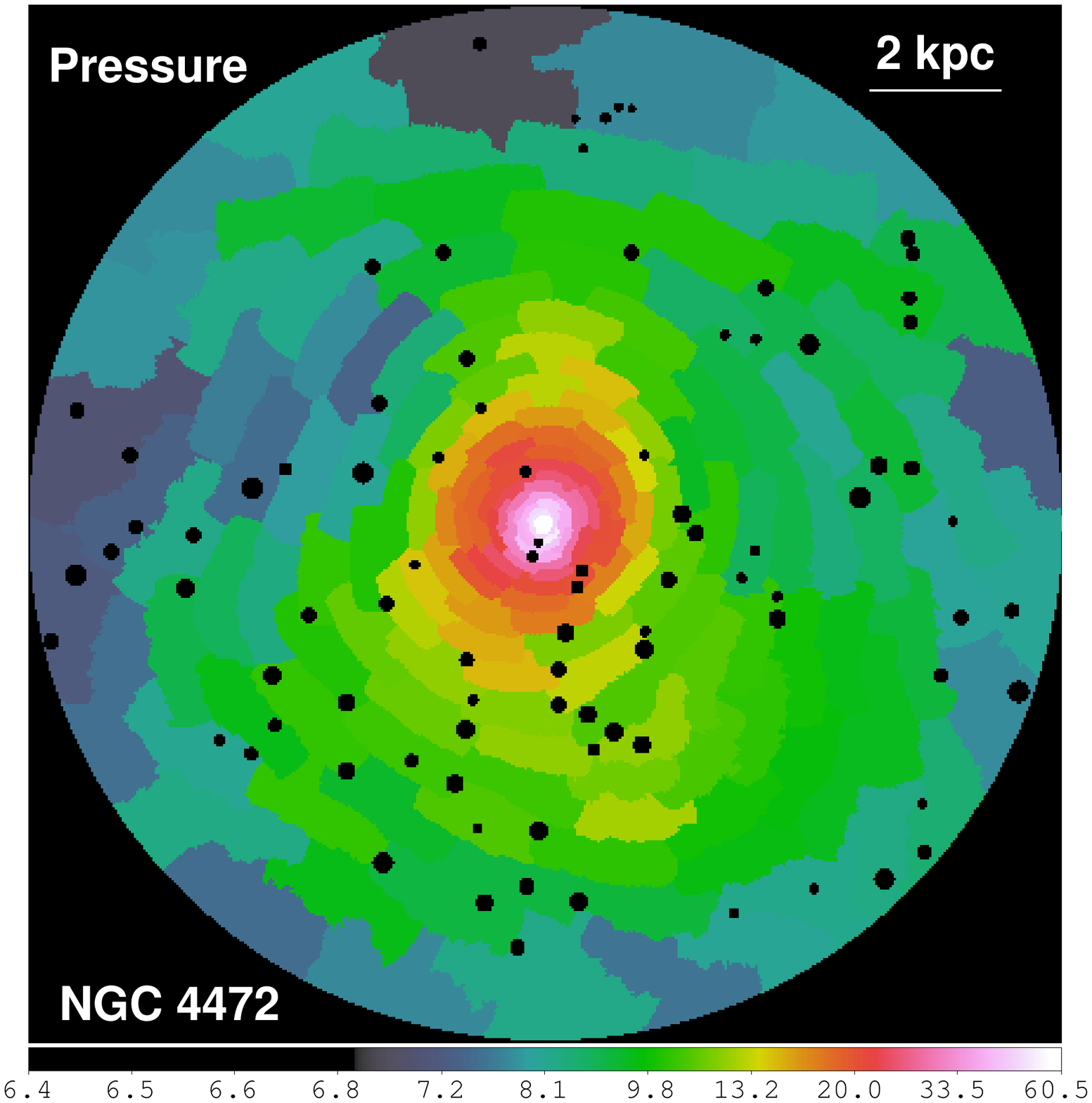}
\end{minipage}
\begin{minipage}{0.32\textwidth}
\includegraphics[height=6cm,clip=t,angle=0.,bb=36 109 577 683]{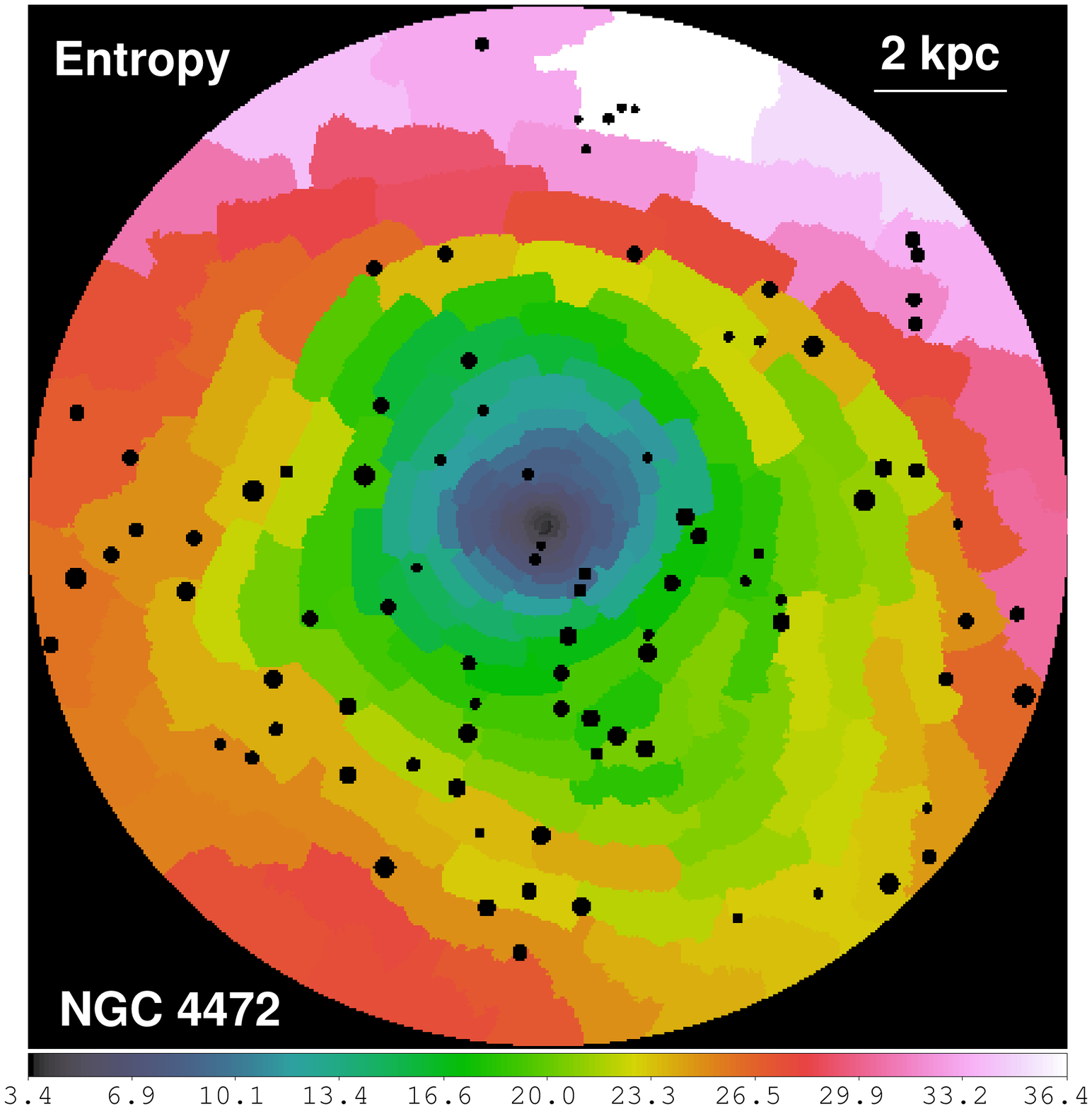}
\end{minipage}
\begin{minipage}{0.32\textwidth}
\includegraphics[height=6cm,clip=t,angle=0.,bb=36 109 577 683]{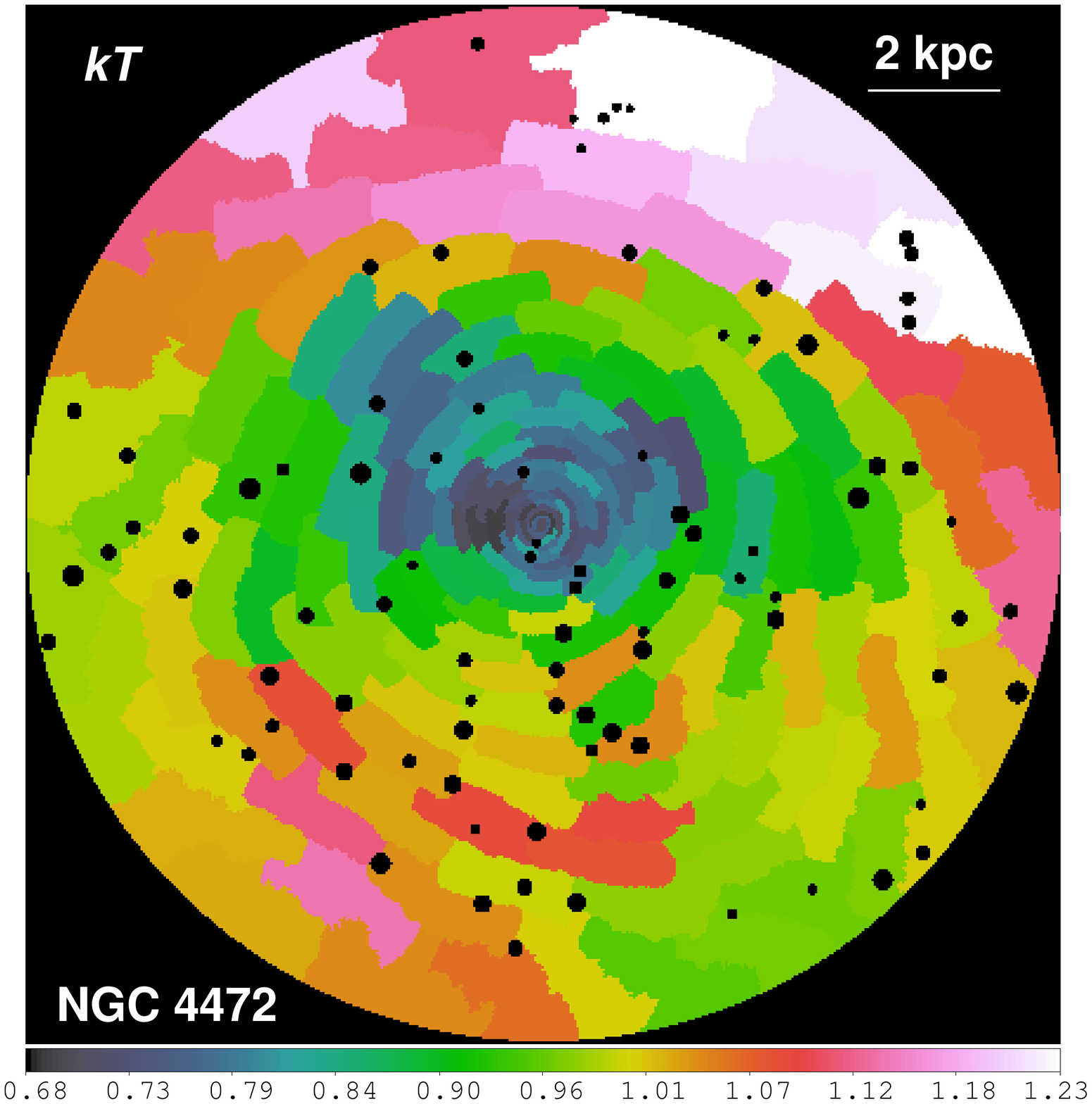}
\end{minipage}
\caption{{\it Top panels:} {\it Chandra} X-ray image of NGC~4472 in the 0.5--2~keV band (left panel); K-band near-infrared image from the Two Micron All-Sky Survey (2MASS; central panel); VLA 1.4 GHz radio map obtained in the C-configuration from \citet{dunn2010} with the X-ray contours over-plotted (right panel). The images are shown in on a logarithmic scale. {\it Bottom panels:} 2D map of pressure (in units of keV~cm$^{-3}\times\left(\frac{l}{\mathrm{20kpc}}\right)^{-1/2}$; left panel), entropy (in units of keV~cm$^2\times\left(\frac{l}{\mathrm{20kpc}}\right)^{1/3}$; central panel), and temperature (in units of keV; right panel).  The maps were obtained by fitting each region independently with a single temperature thermal model, yielding 1$\sigma$ fractional uncertainties of $\sim$3--5 per cent in temperature, entropy, and pressure. Point sources were excluded from the spectral analysis and therefore appear as black circles on the maps. }
\label{N4472}
\end{figure*}

\begin{figure*}
\begin{minipage}{0.32\textwidth}
\includegraphics[height=6cm,clip=t,angle=0.,bb=36 109 577 683]{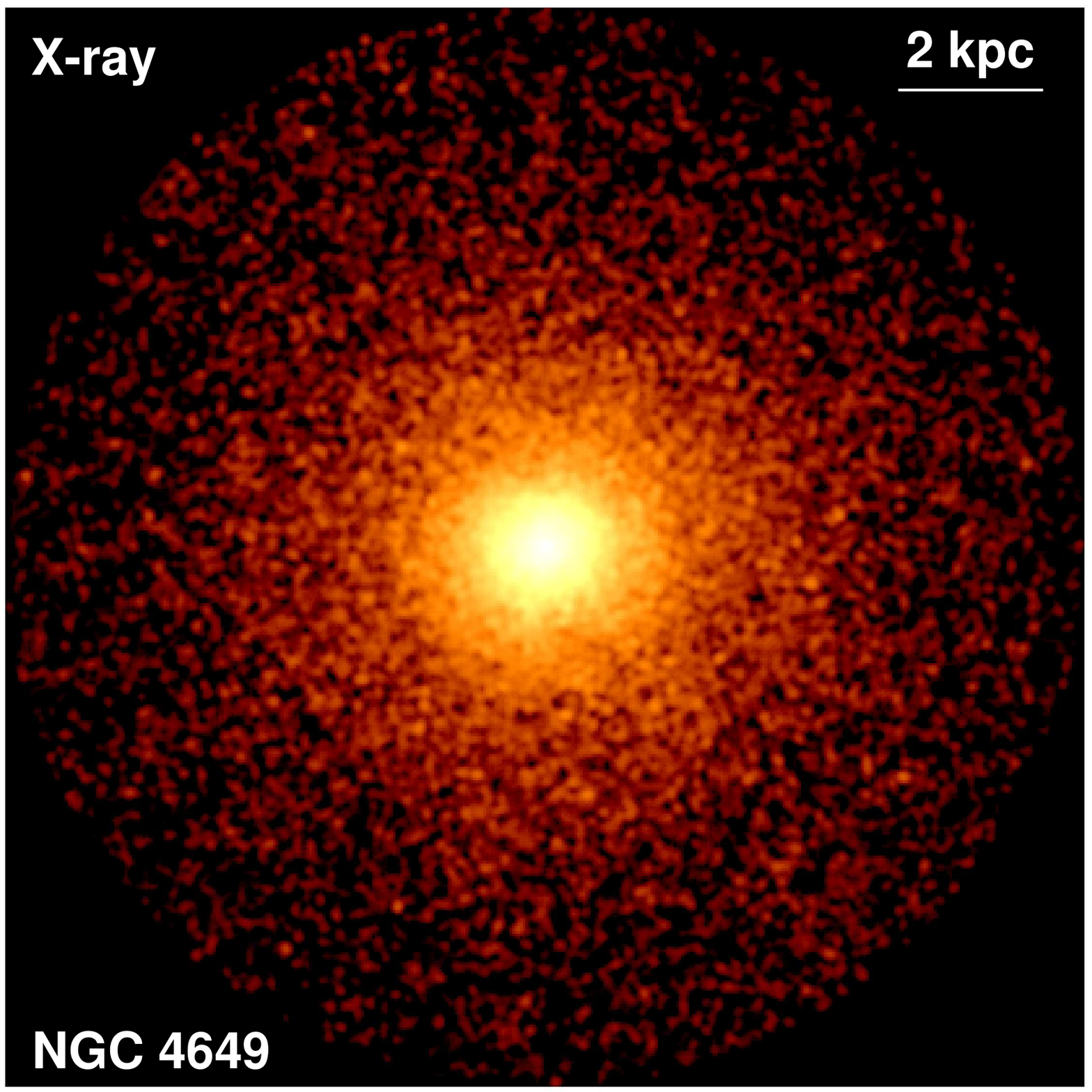}
\end{minipage} 
\begin{minipage}{0.32\textwidth}
\includegraphics[height=6cm,clip=t,angle=0.,bb=36 109 577 683]{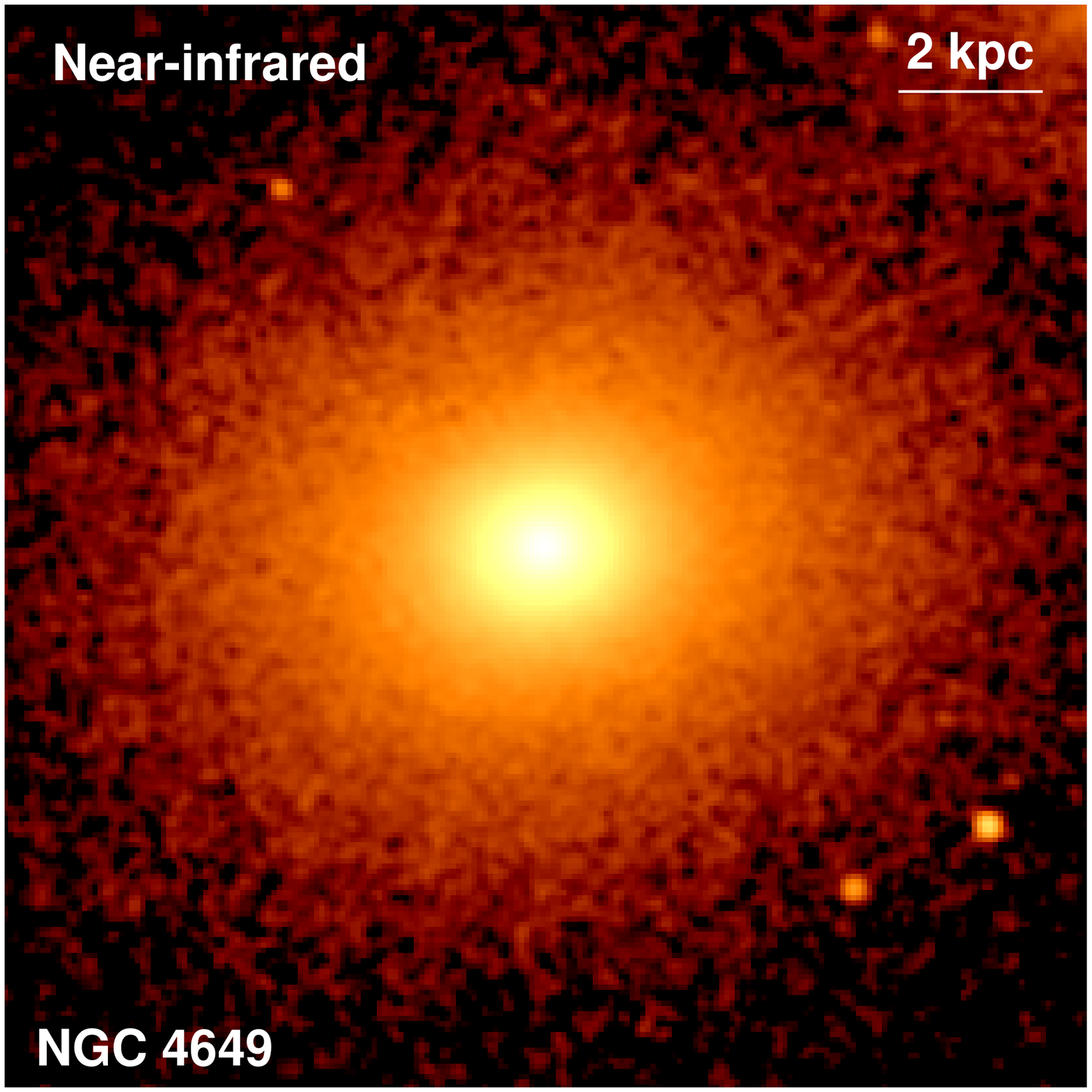}
\end{minipage} 
\begin{minipage}{0.32\textwidth}
\includegraphics[height=6cm,clip=t,angle=0.,bb=36 109 577 683]{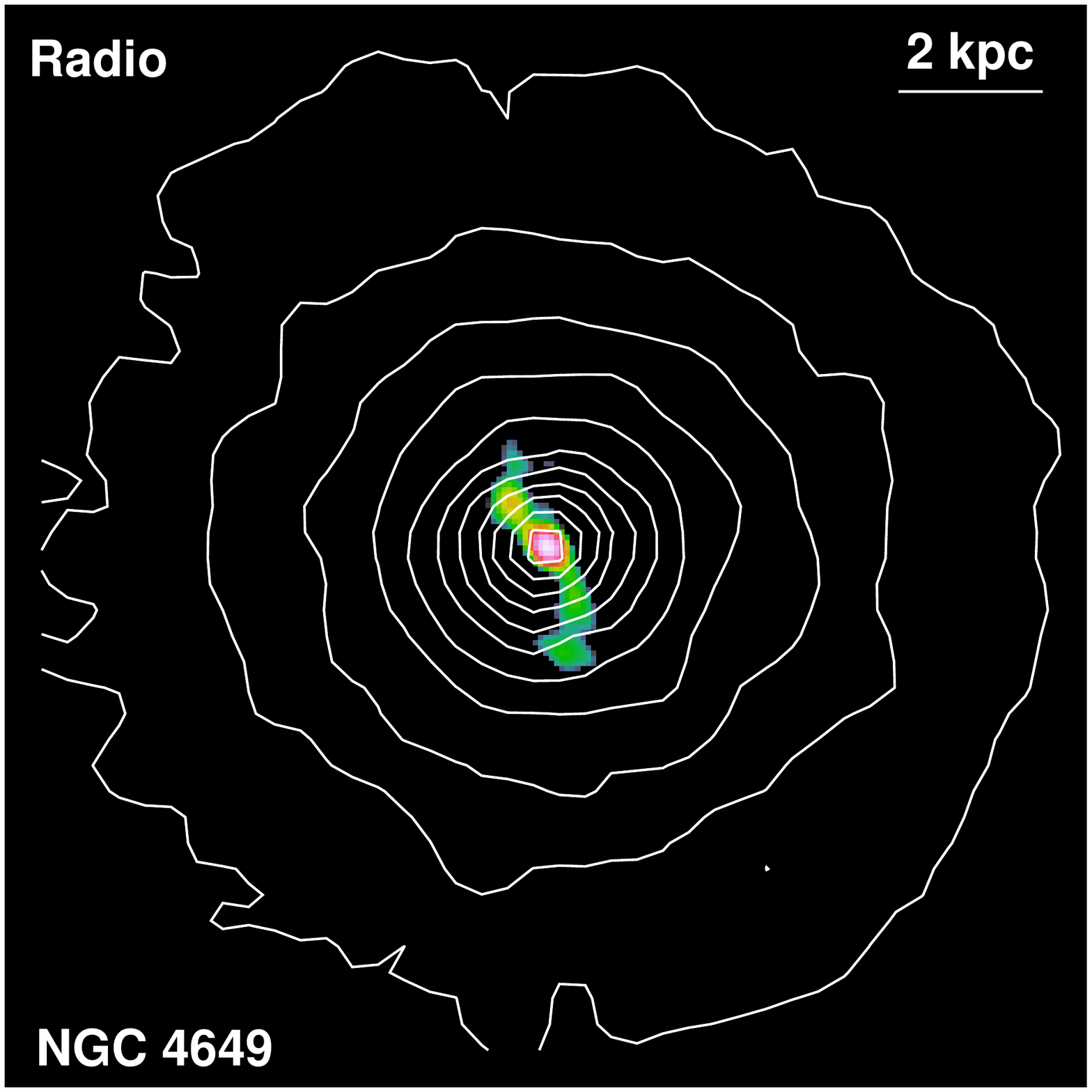}
\end{minipage} 
\begin{minipage}{0.32\textwidth}
\includegraphics[height=6cm,clip=t,angle=0.,bb=36 109 577 683]{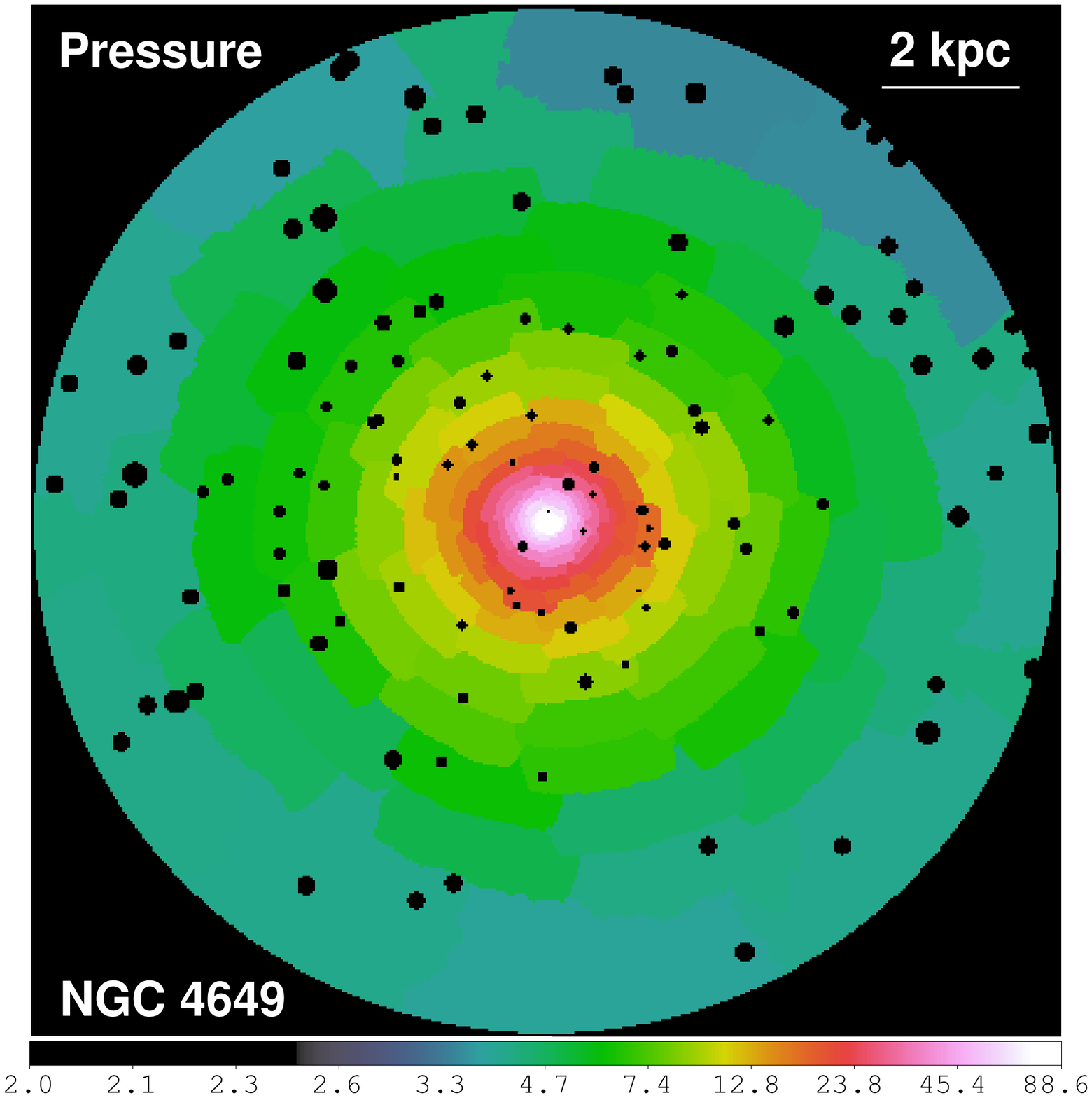}
\end{minipage}
\begin{minipage}{0.32\textwidth}
\includegraphics[height=6cm,clip=t,angle=0.,bb=36 109 577 683]{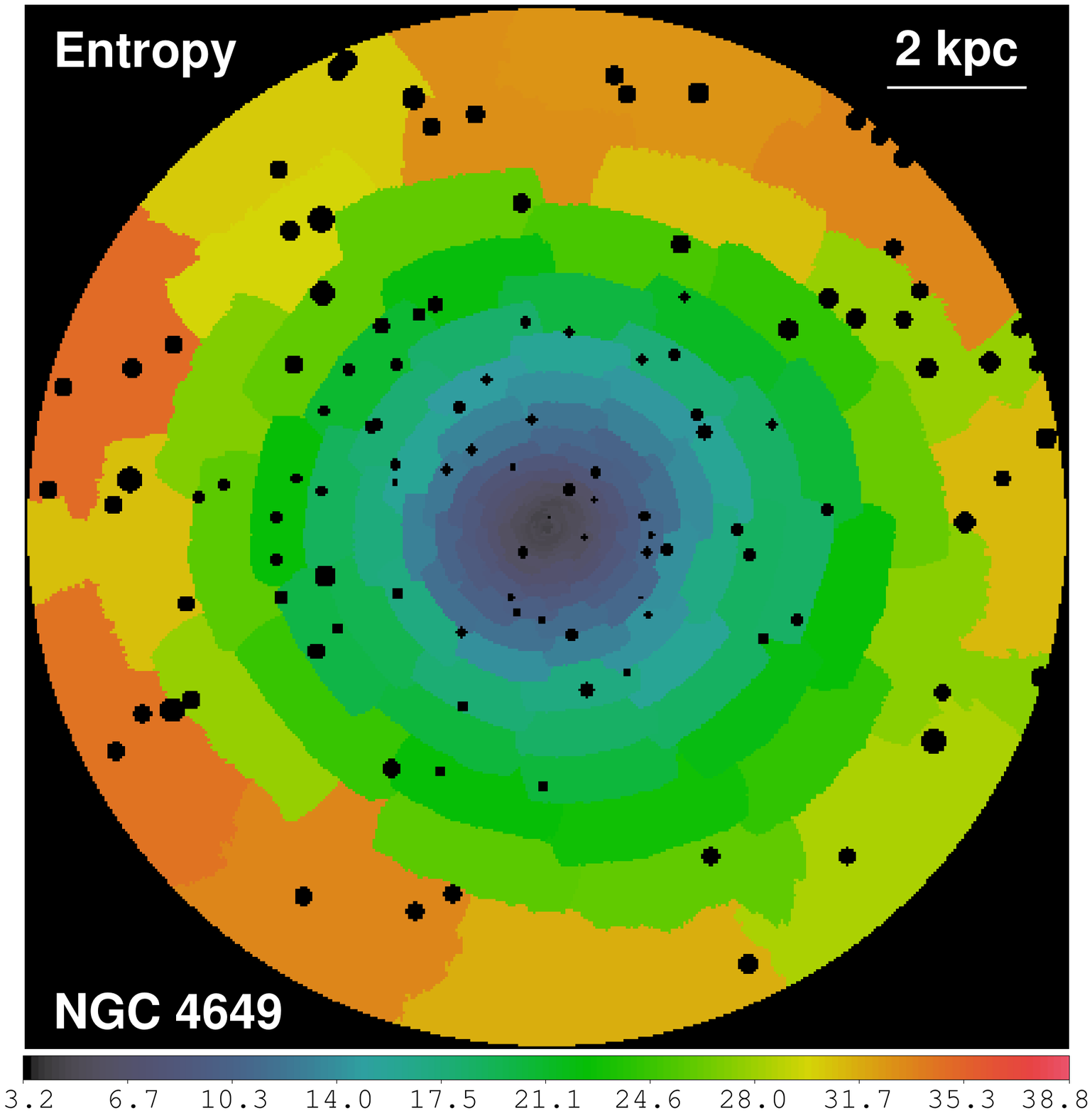}
\end{minipage}
\begin{minipage}{0.32\textwidth}
\includegraphics[height=6cm,clip=t,angle=0.,bb=36 109 577 683]{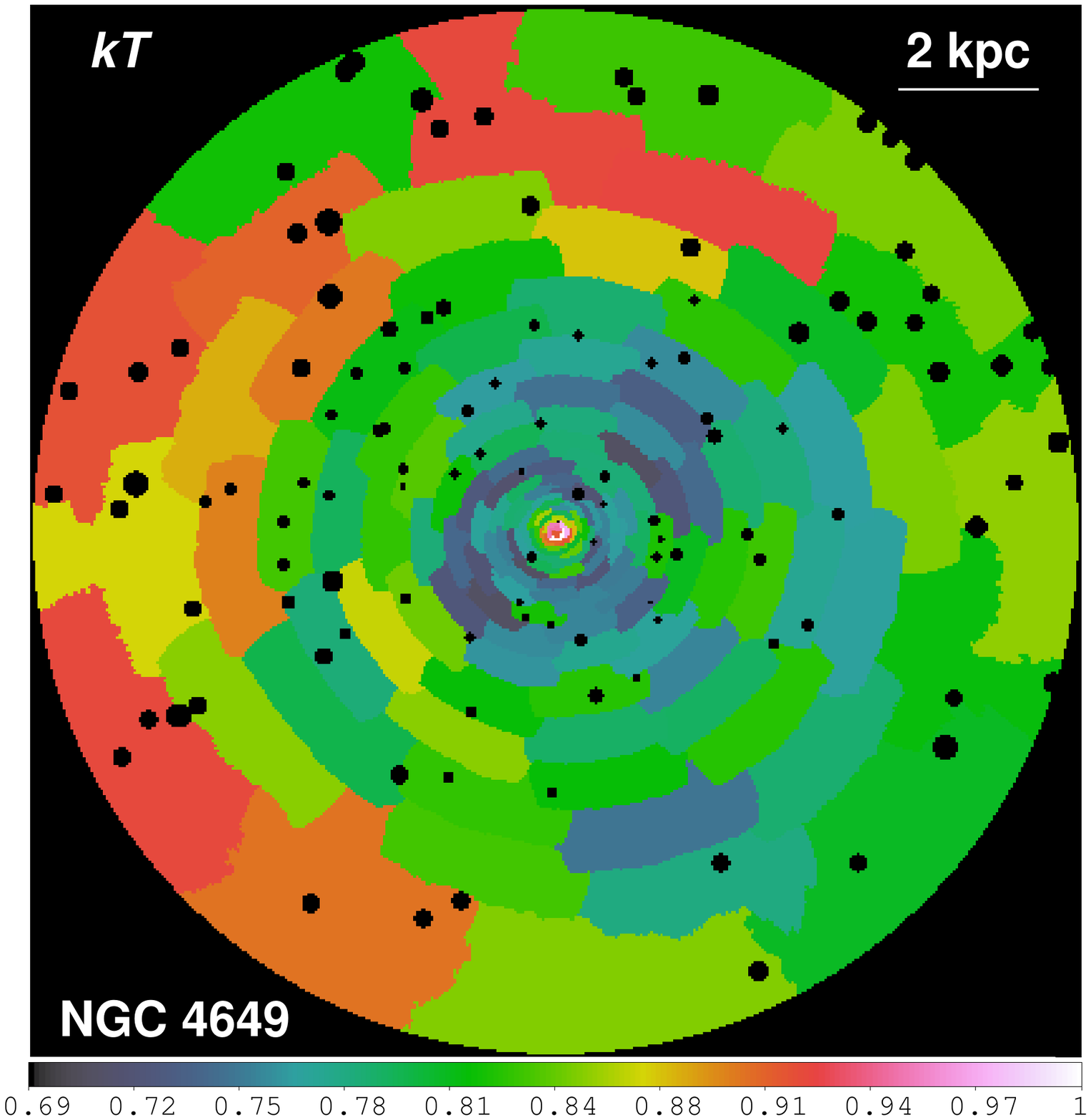}
\end{minipage}
\caption{As Fig. \ref{N4472}, but for NGC~4649. The VLA 1.4 GHz radio map was obtained in the B-configuration \citep{dunn2010}.
}
\label{N4649}
\end{figure*}

\begin{figure*}
\begin{minipage}{0.32\textwidth}
\includegraphics[height=6cm,clip=t,angle=0.,bb=36 109 577 683]{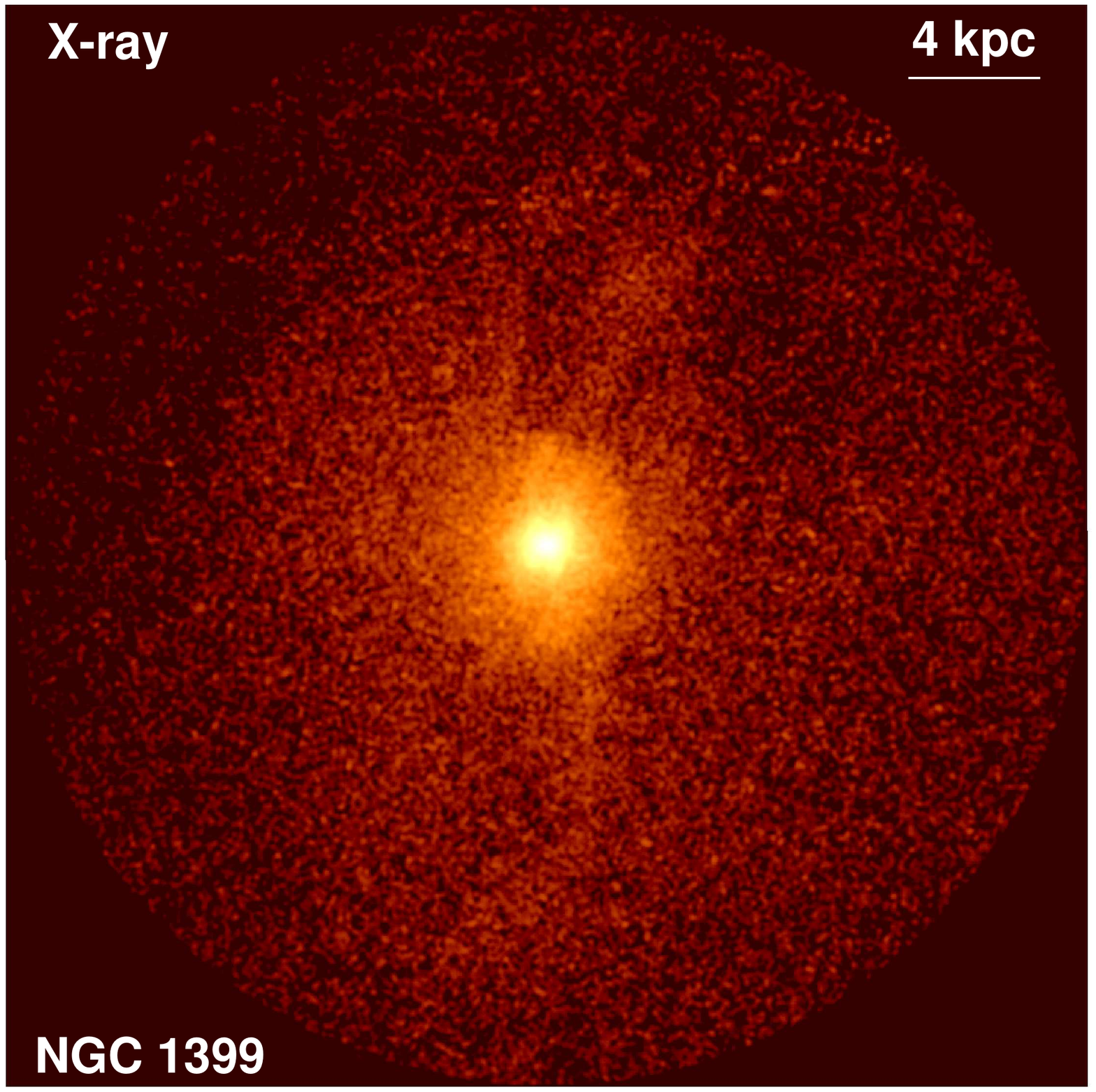}
\end{minipage} 
\begin{minipage}{0.32\textwidth}
\includegraphics[height=6cm,clip=t,angle=0.,bb=36 109 577 683]{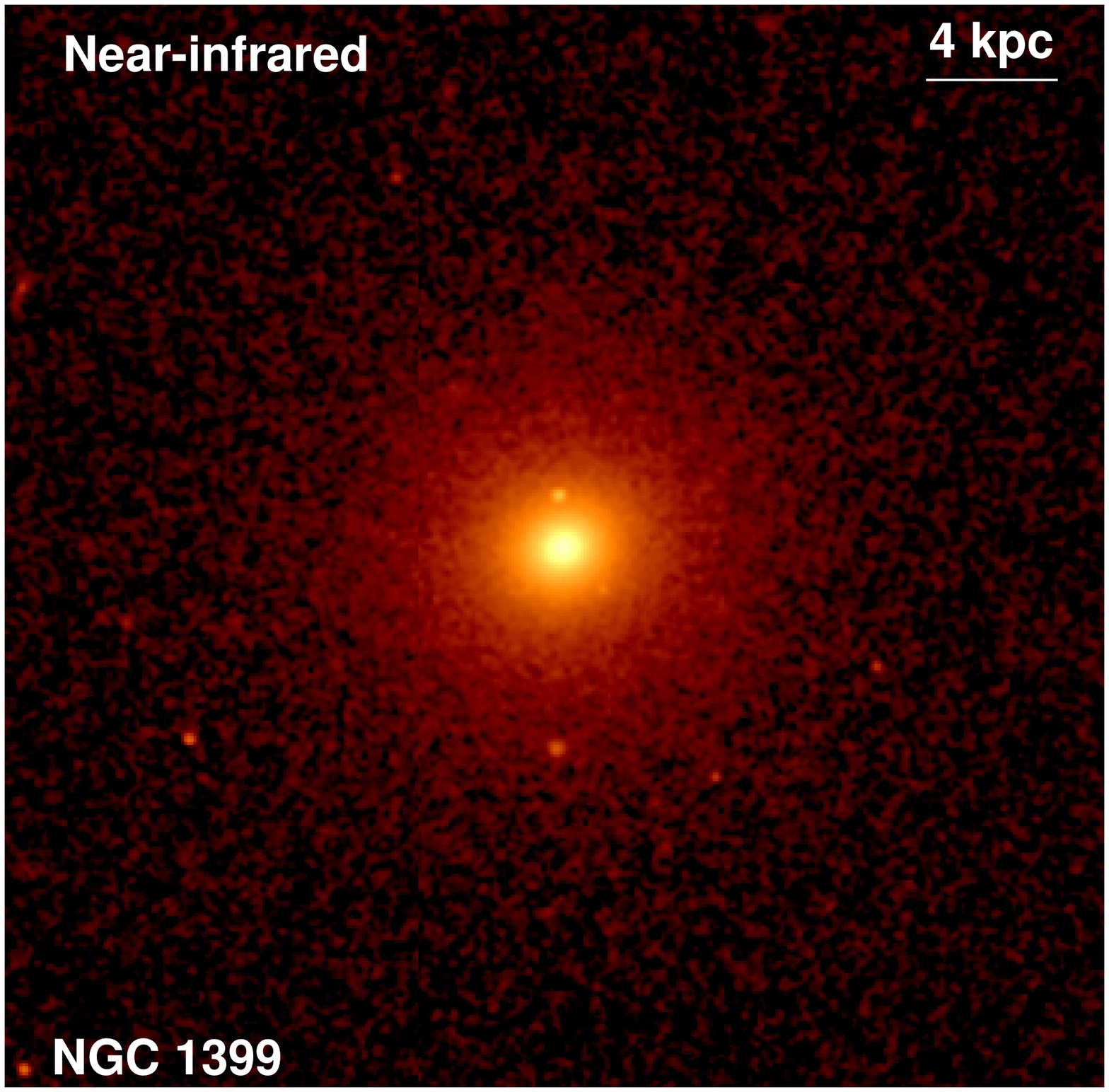}
\end{minipage} 
\begin{minipage}{0.32\textwidth}
\includegraphics[height=6cm,clip=t,angle=0.,bb=36 109 577 683]{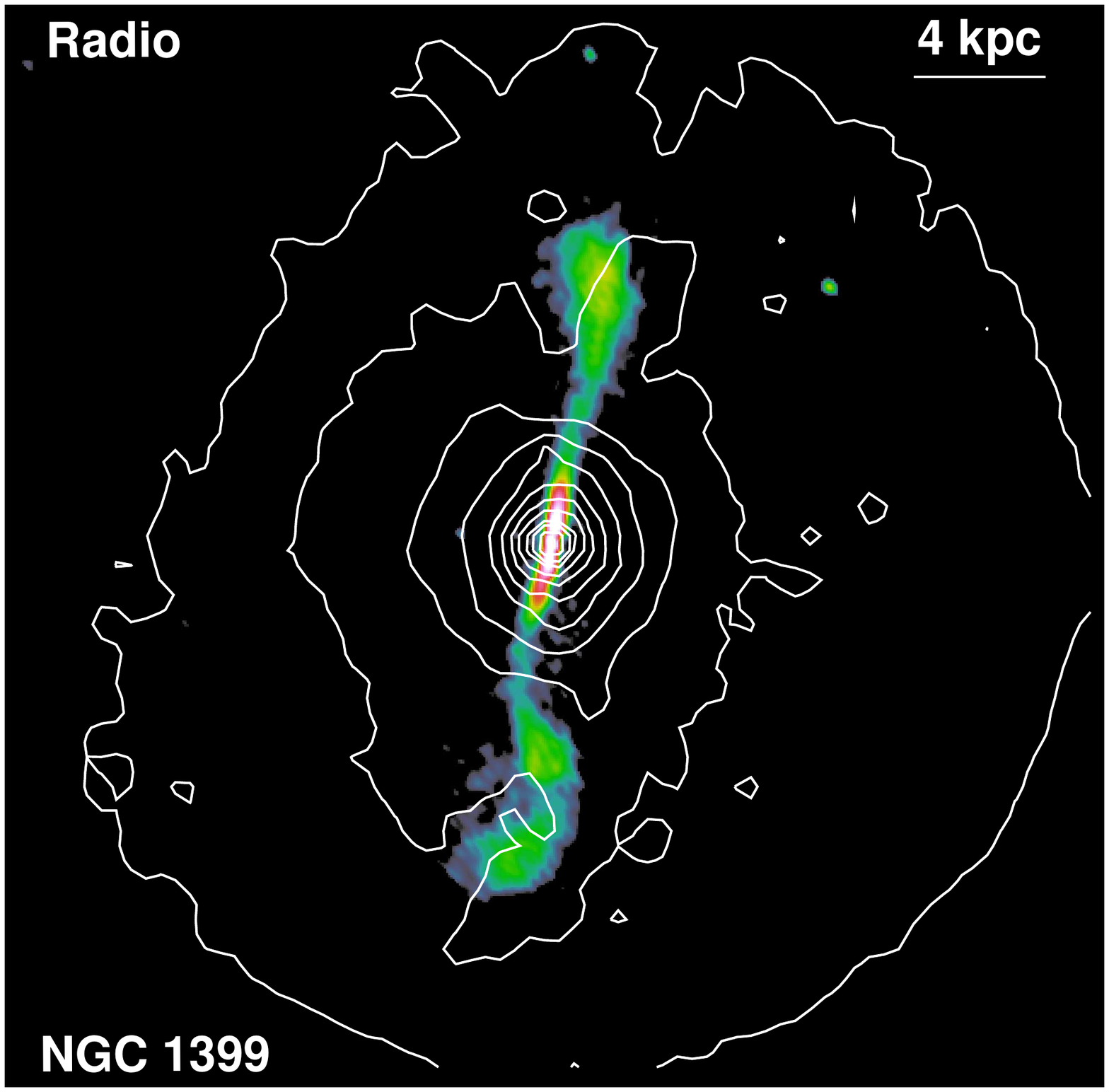}
\end{minipage} 
\begin{minipage}{0.32\textwidth}
\includegraphics[height=6cm,clip=t,angle=0.,bb=36 109 577 683]{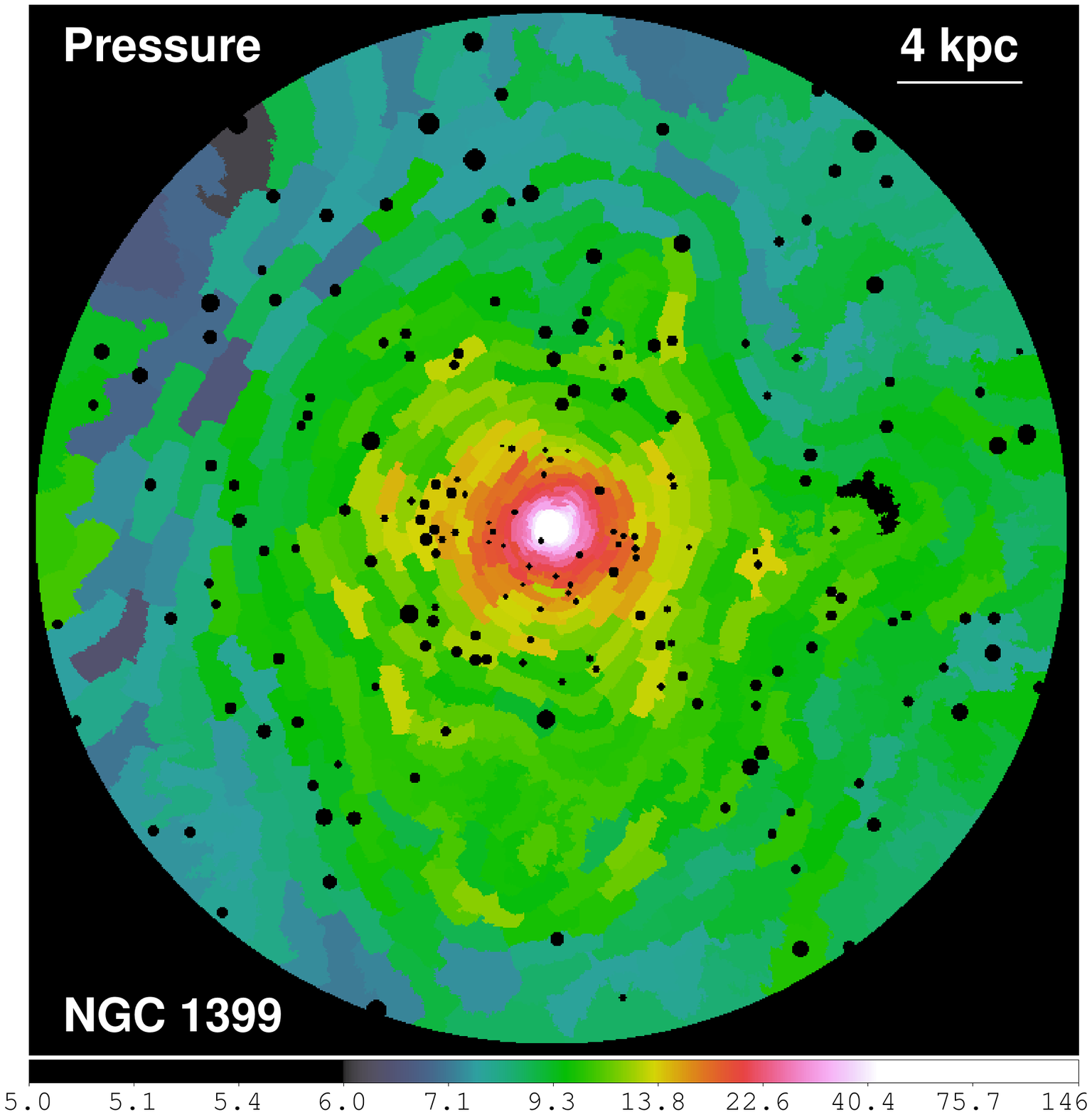}
\end{minipage}
\begin{minipage}{0.32\textwidth}
\includegraphics[height=6cm,clip=t,angle=0.,bb=36 109 577 683]{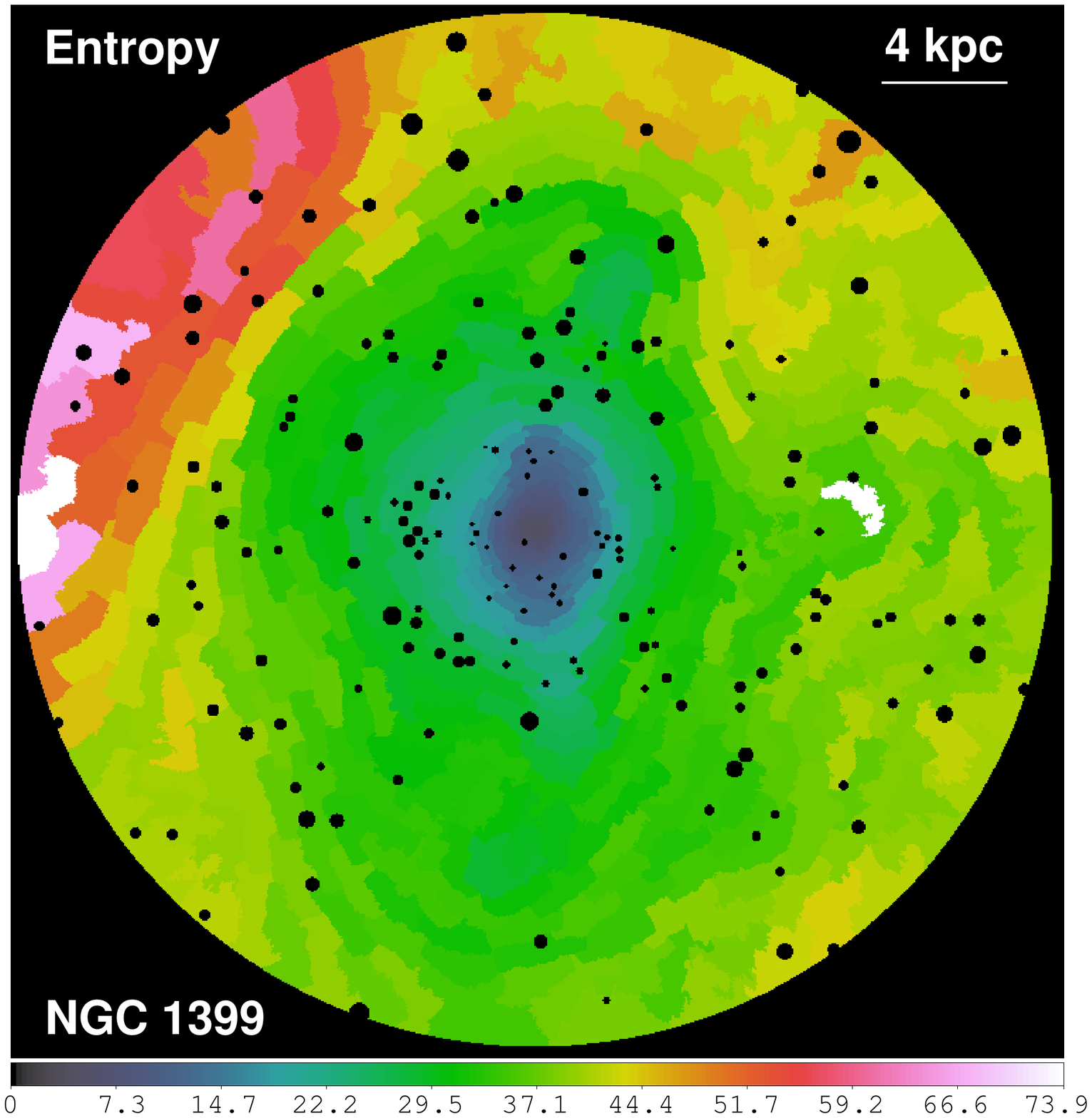}
\end{minipage}
\begin{minipage}{0.32\textwidth}
\includegraphics[height=6cm,clip=t,angle=0.,bb=36 109 577 683]{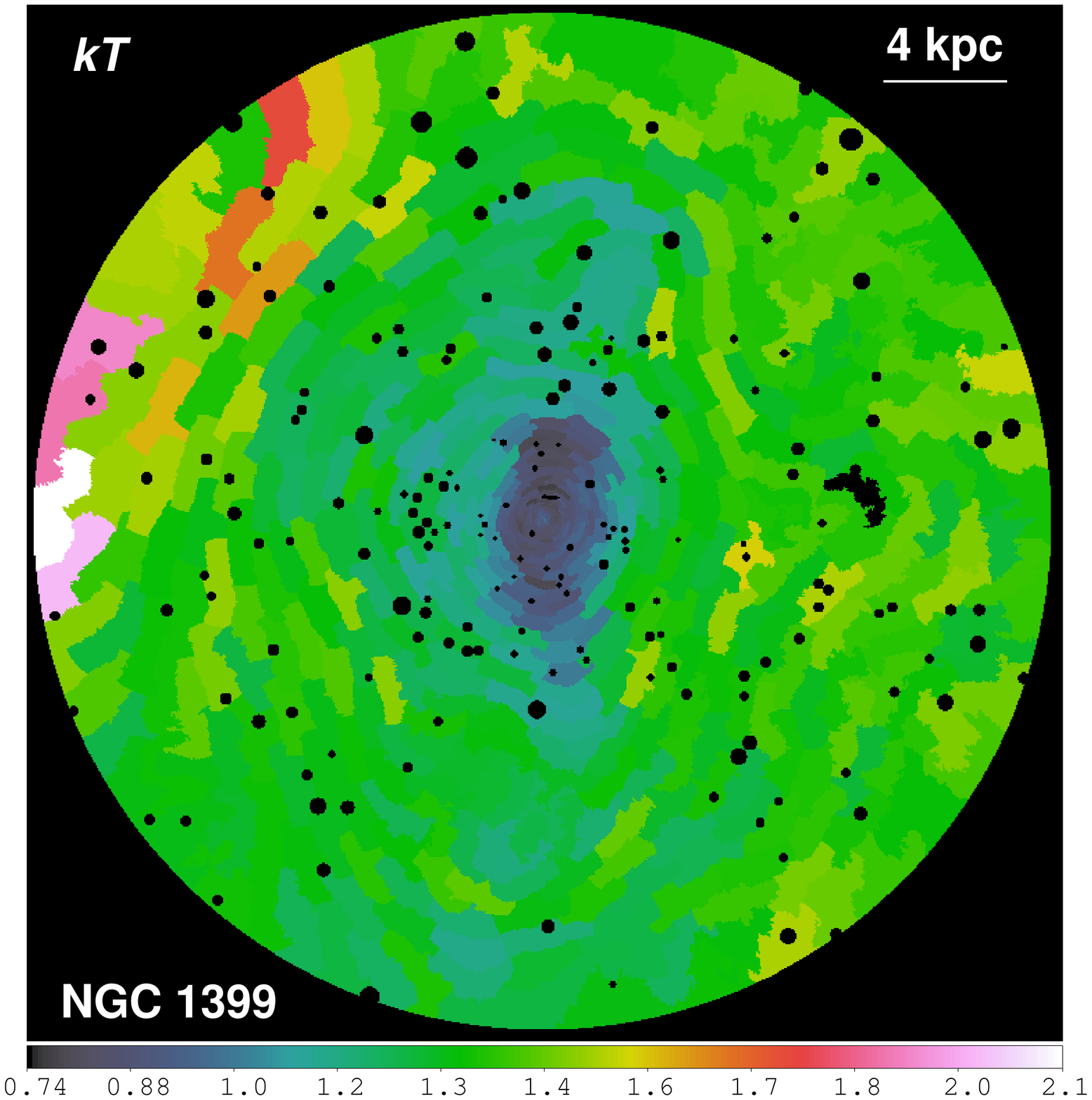}
\end{minipage}
\caption{As Fig. \ref{N4472}, but for NGC~1399. The VLA 1.4 GHz radio map was obtained in the H-configuration \citep[hybrid A and B configuration;][]{dunn2010}.
 }
\label{N1399}
\end{figure*}

\begin{figure*}
\begin{minipage}{0.32\textwidth}
\includegraphics[height=5.5cm,clip=t,angle=0.,bb= 36 117 577 676]{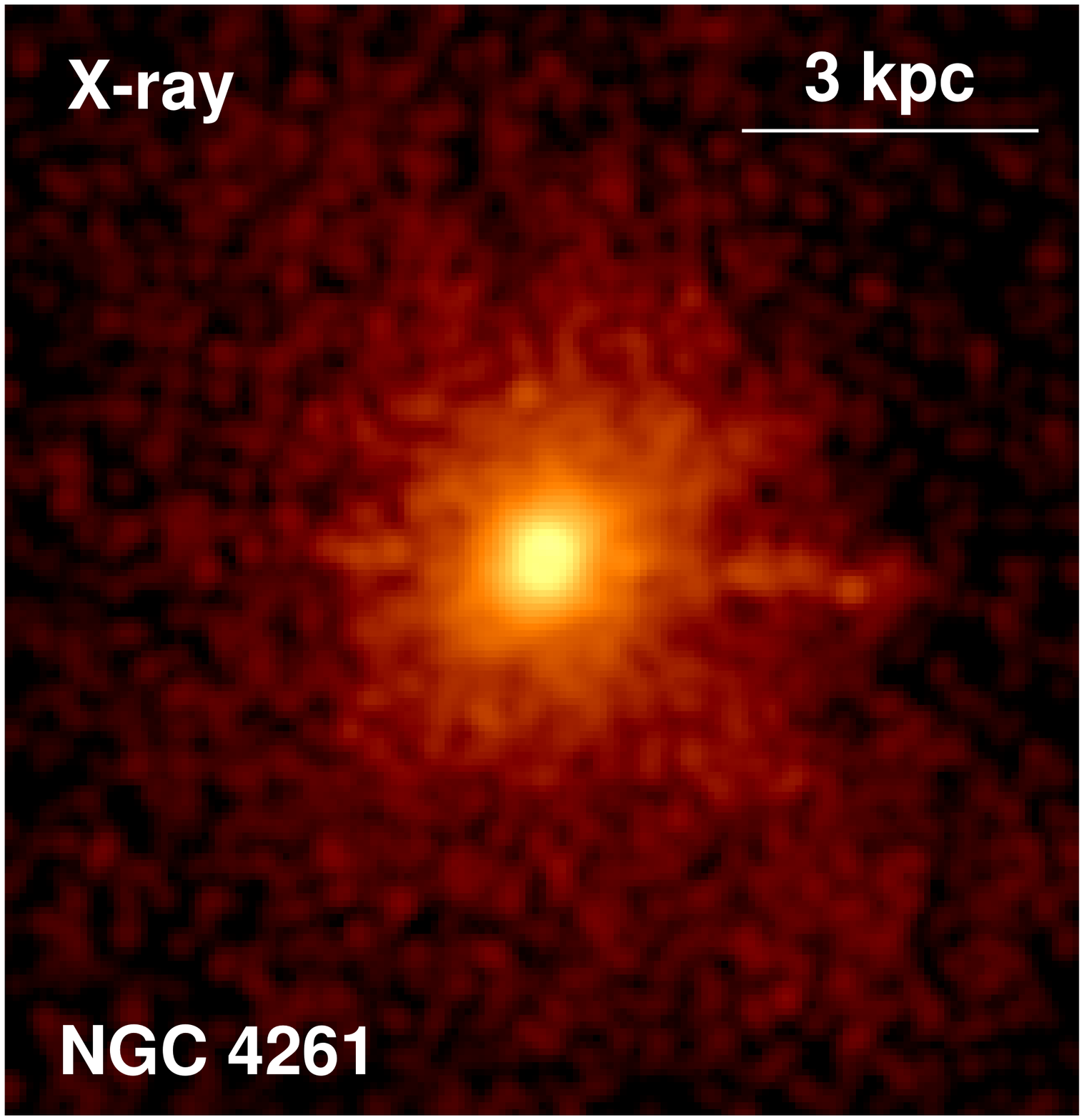}
\end{minipage} 
\begin{minipage}{0.32\textwidth}
\includegraphics[height=5.5cm,clip=t,angle=0.,bb=36 117 577 676]{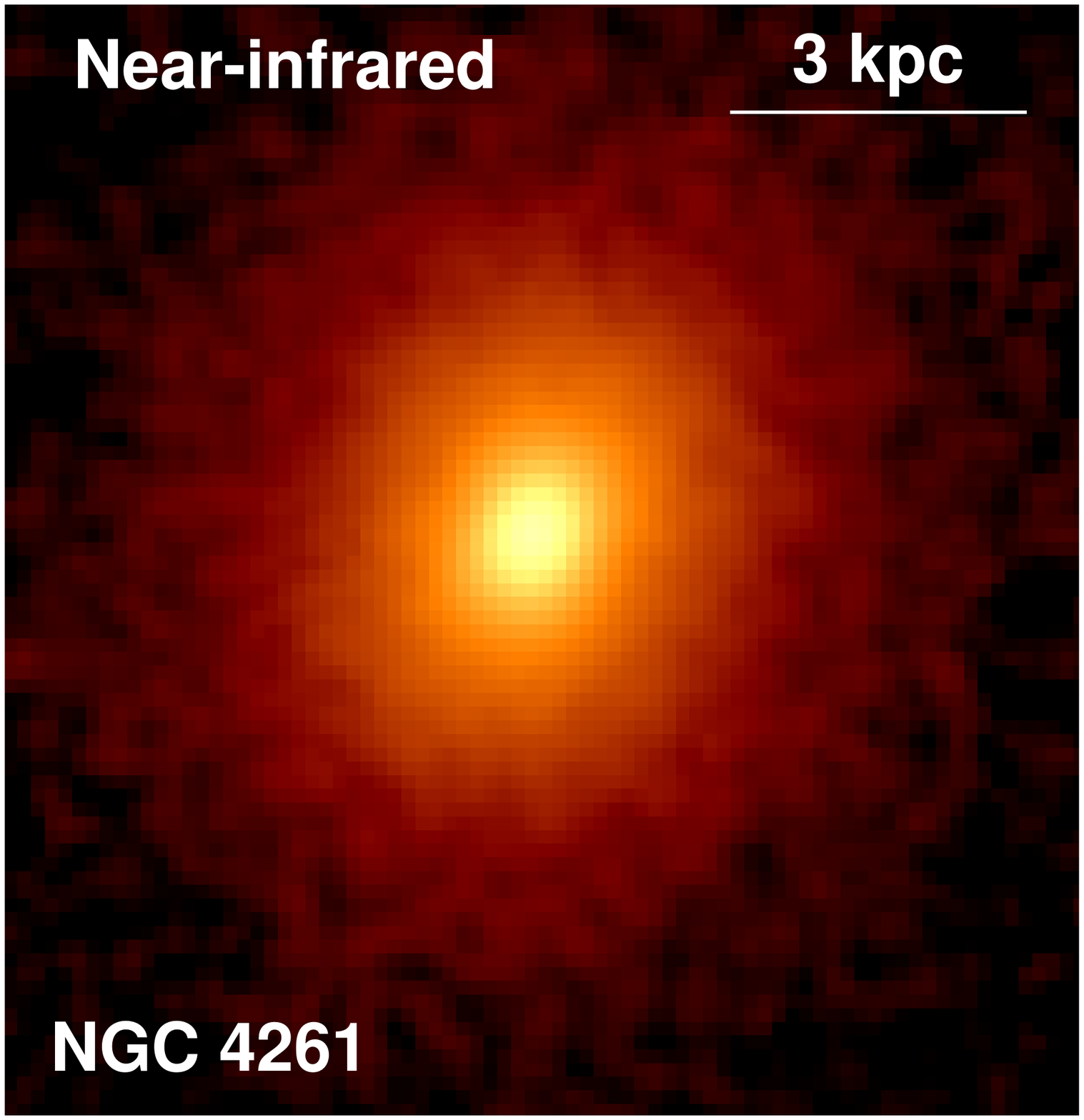}
\end{minipage} 
\begin{minipage}{0.32\textwidth}
\includegraphics[height=5.5cm,width=5.7cm,clip=t,angle=0.,bb=36 151 577 642]{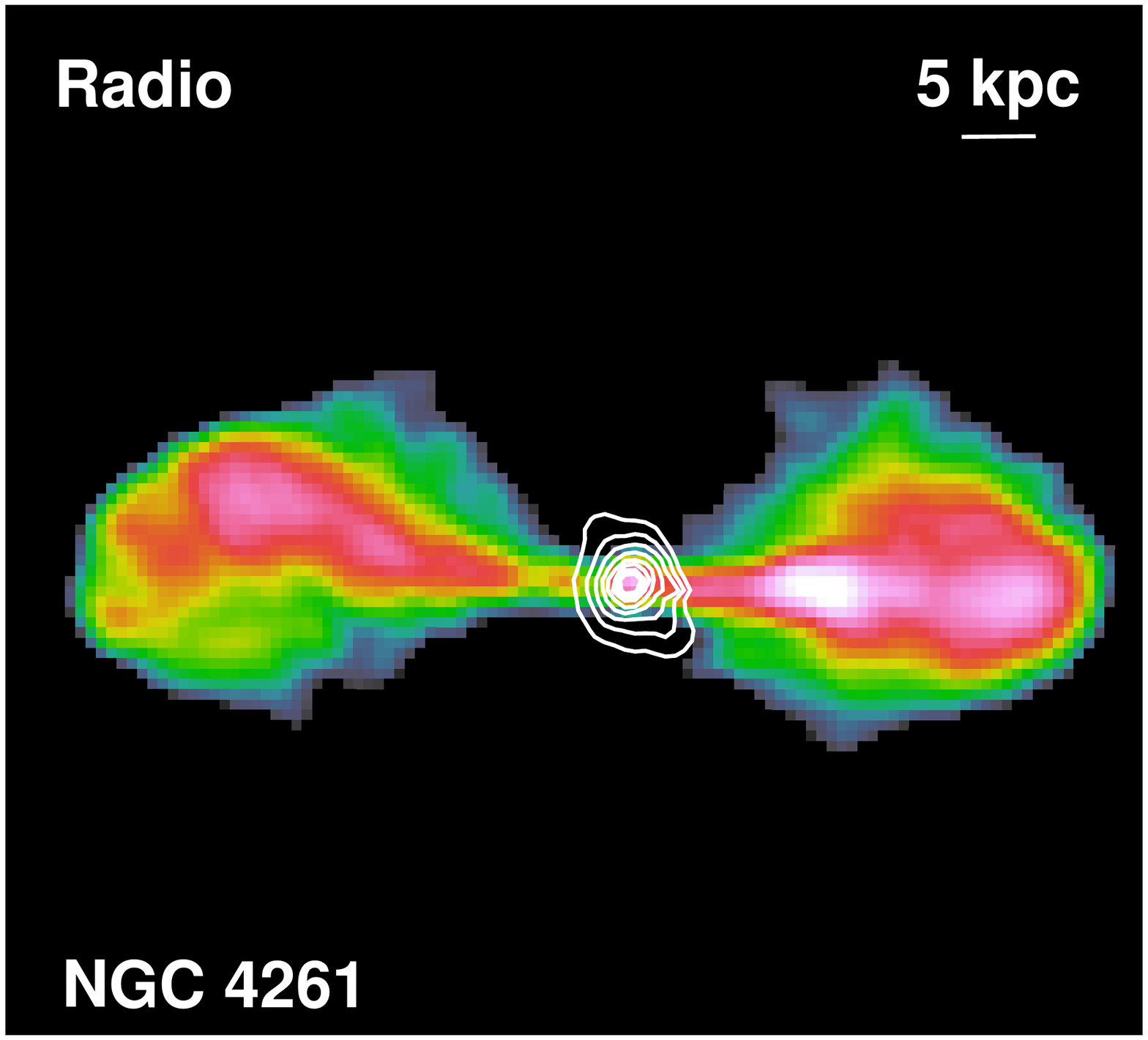}
\end{minipage} \\
\vspace{0.5cm}
\begin{minipage}{0.32\textwidth}
\includegraphics[height=6cm,clip=t,angle=0.,bb=36 109 577 683]{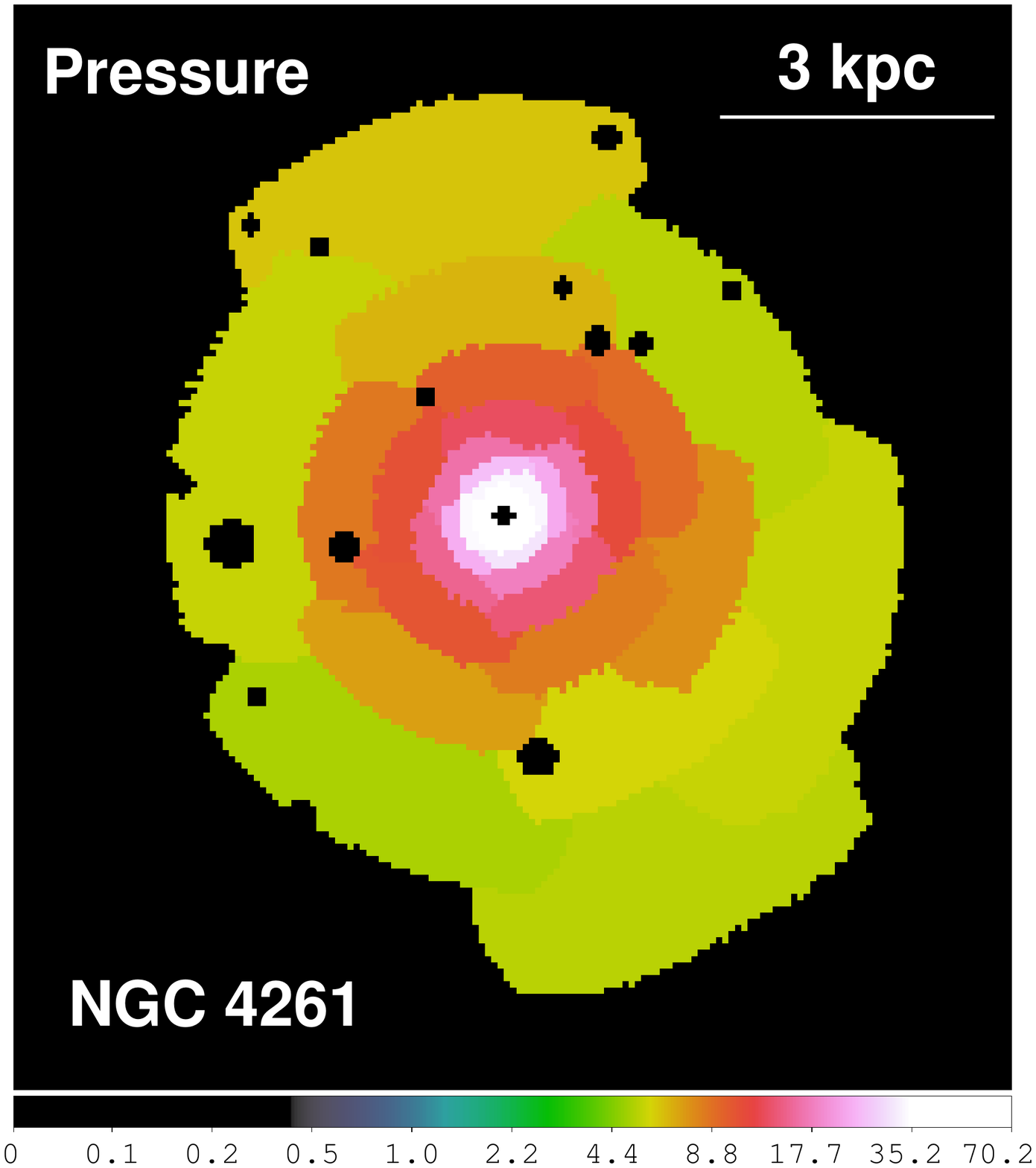}
\end{minipage}
\begin{minipage}{0.32\textwidth}
\includegraphics[height=6cm,clip=t,angle=0.,bb=36 109 577 683]{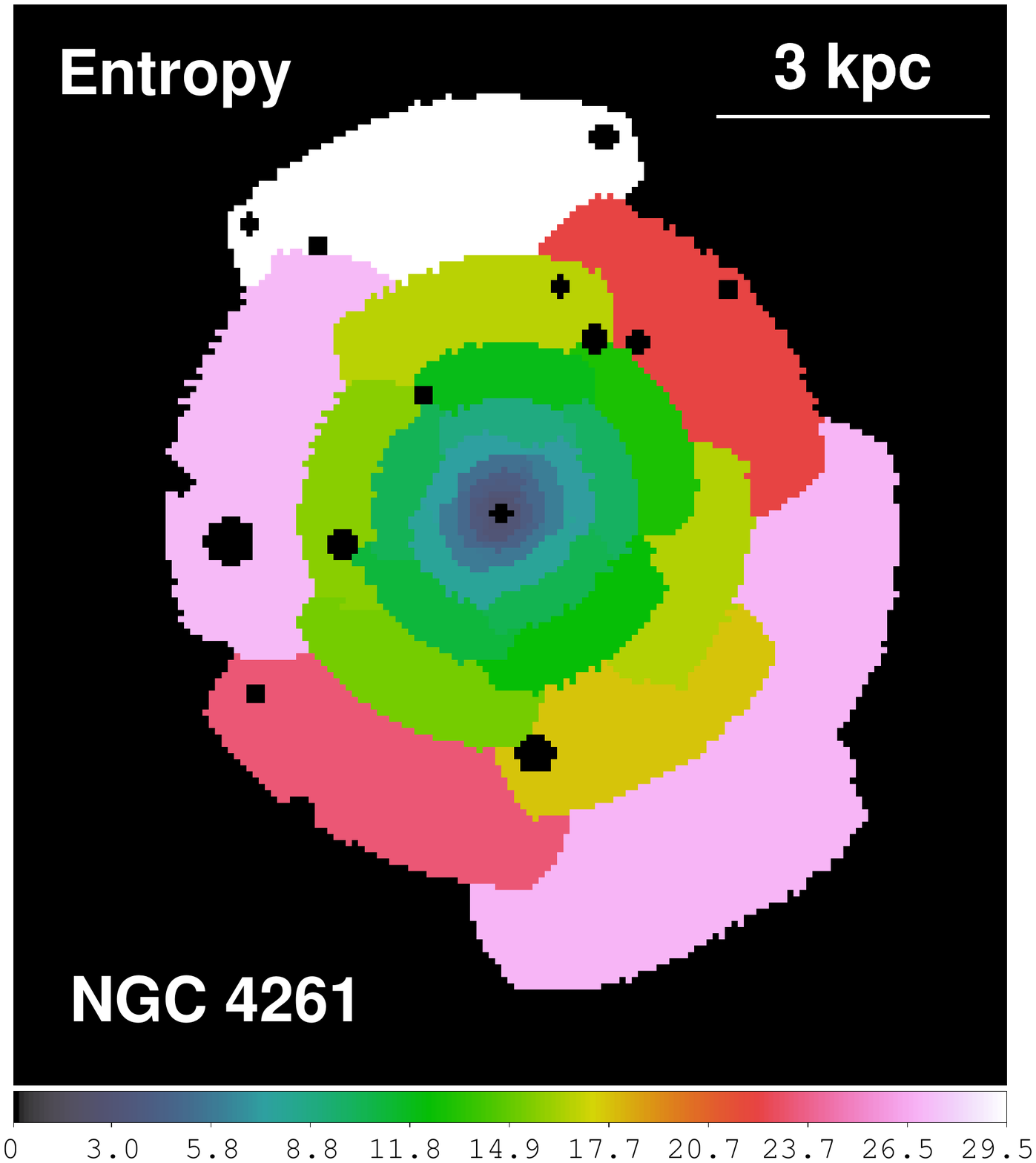}
\end{minipage}
\begin{minipage}{0.32\textwidth}
\includegraphics[height=6cm,clip=t,angle=0.,bb=36 109 577 683]{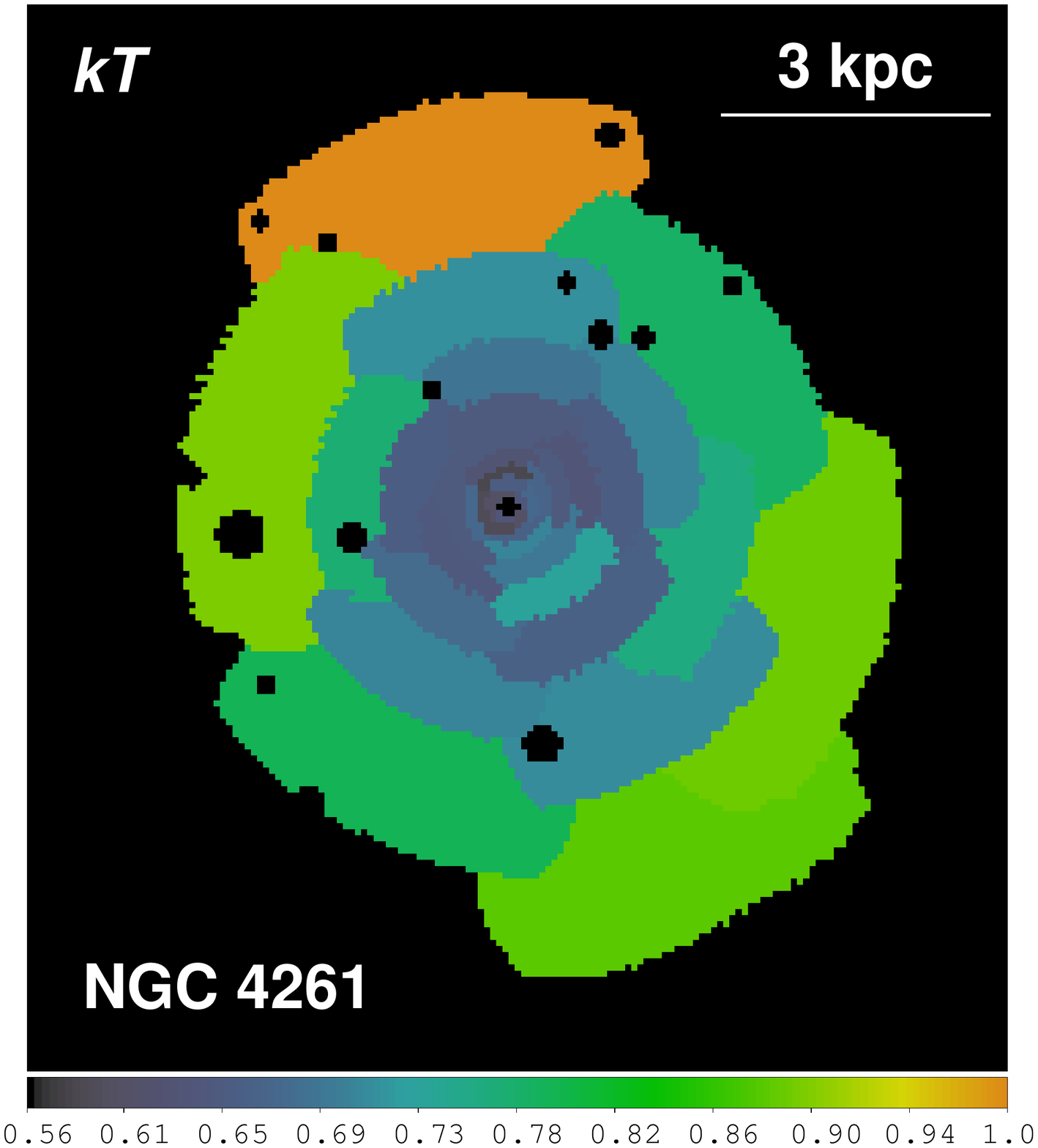}
\end{minipage}
\caption{As Fig. \ref{N4472}, but for NGC~4261. }
\label{N4261}
\end{figure*}

\begin{figure*}
\begin{minipage}{0.45\textwidth}
\hspace{2cm}\includegraphics[height=6cm,clip=t,angle=0.,bb=36 123 577 669]{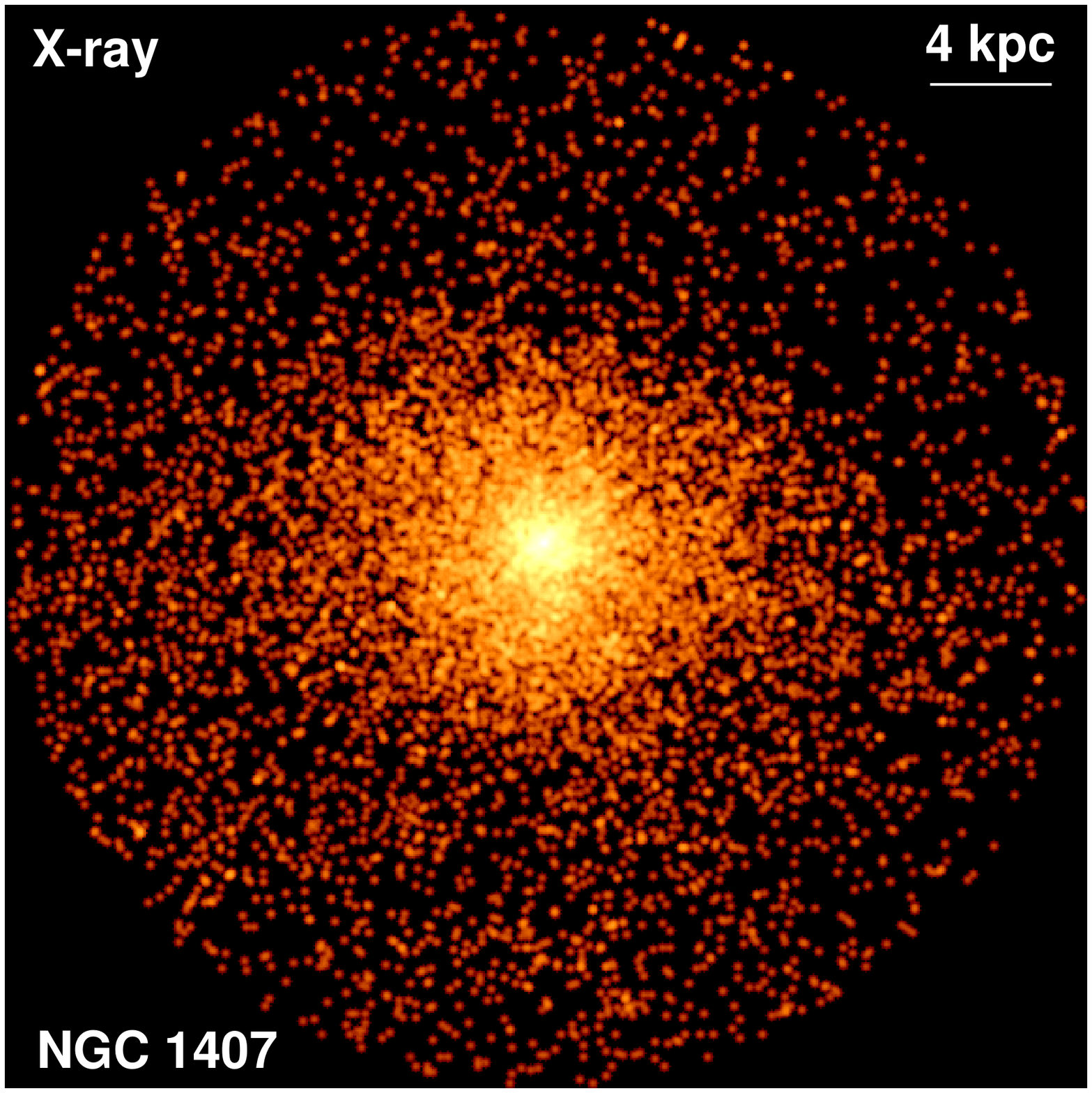}
\end{minipage}
\begin{minipage}{0.45\textwidth}
\includegraphics[height=5.75cm,clip=t,angle=0.,bb=36 123 577 669]{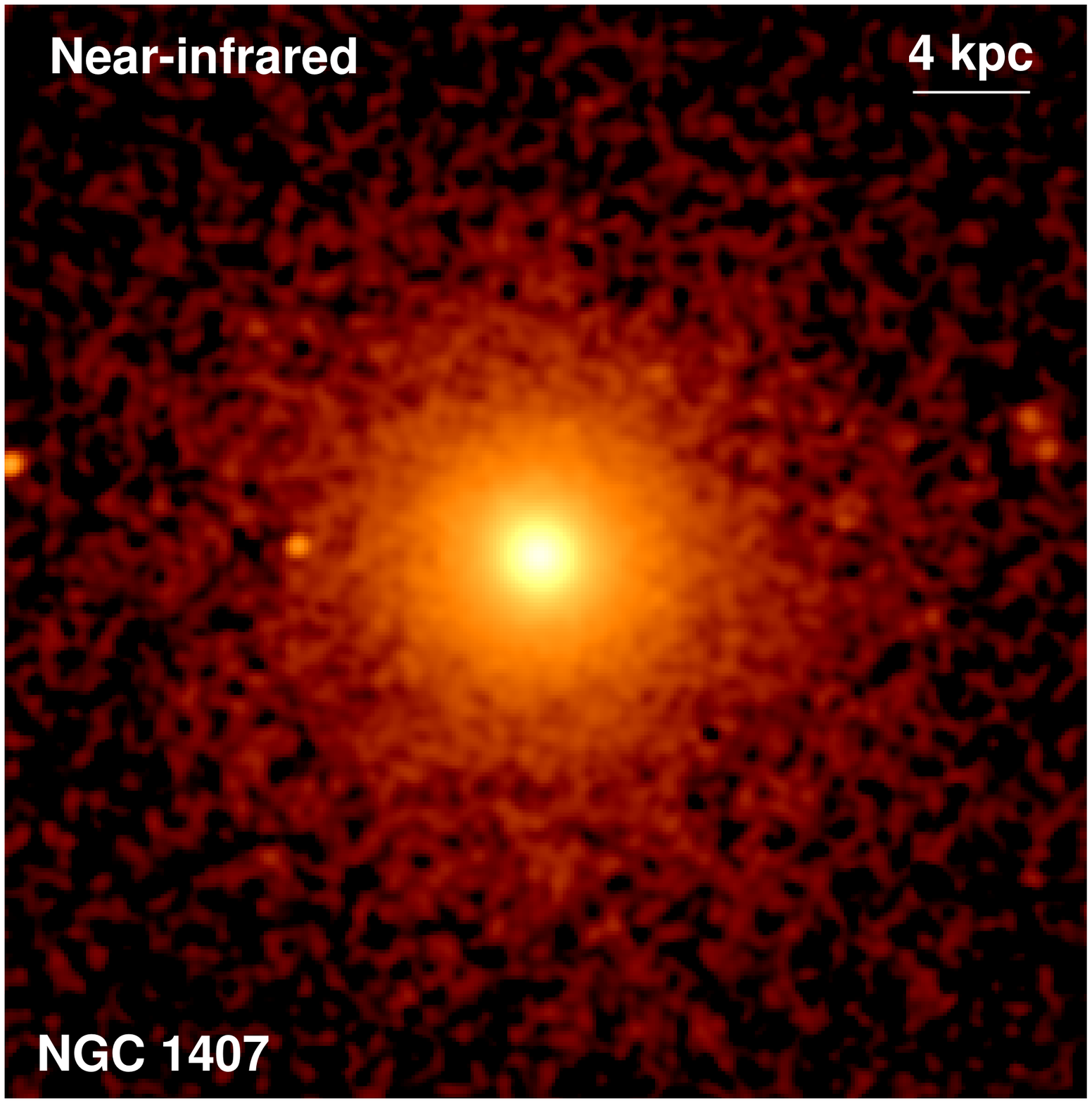}
\end{minipage}\\
\begin{minipage}{0.32\textwidth}
\includegraphics[height=6cm,clip=t,angle=0.,bb=36 109 577 683]{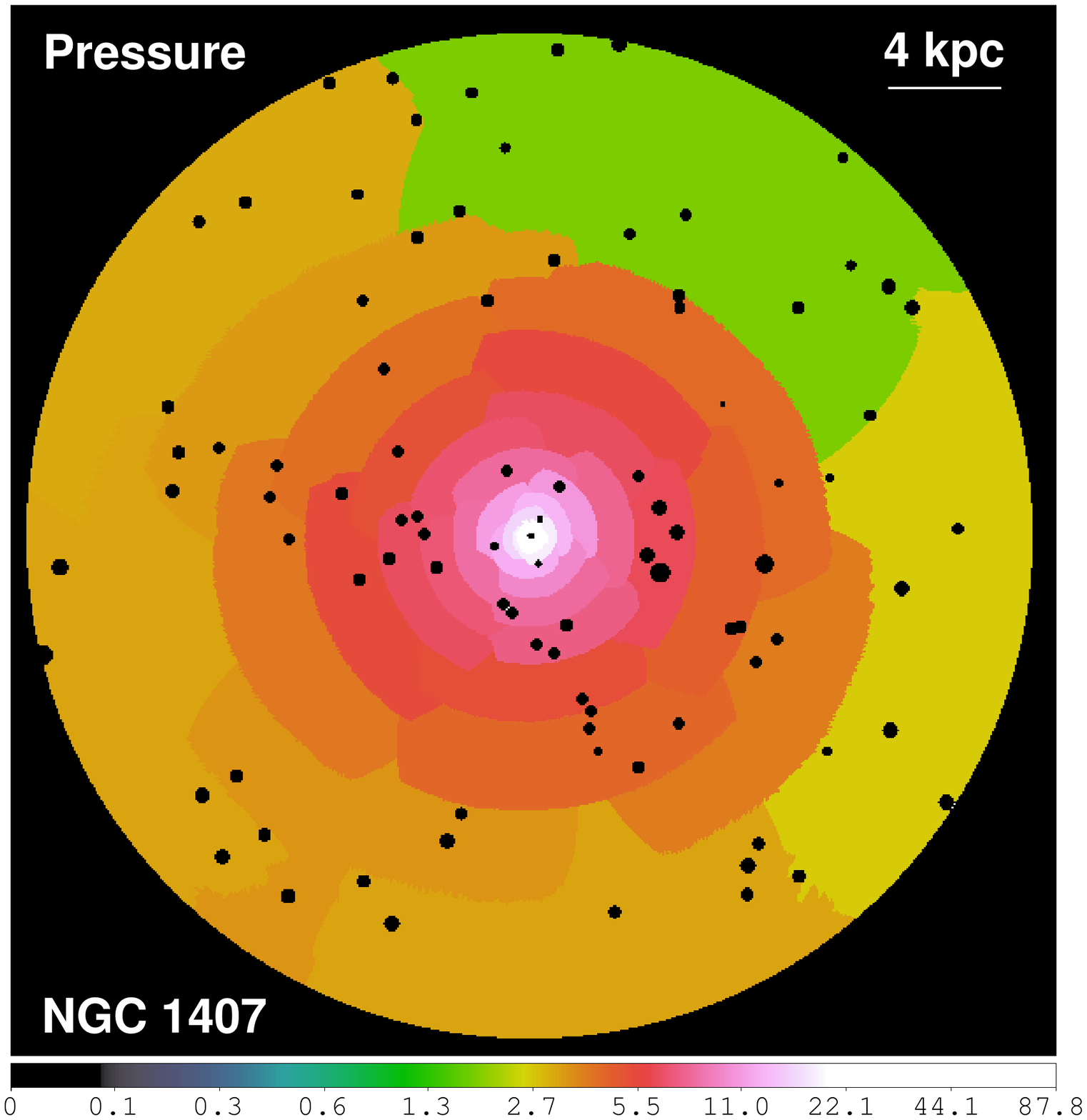}
\end{minipage}
\begin{minipage}{0.32\textwidth}
\includegraphics[height=6cm,clip=t,angle=0.,bb=36 109 577 683]{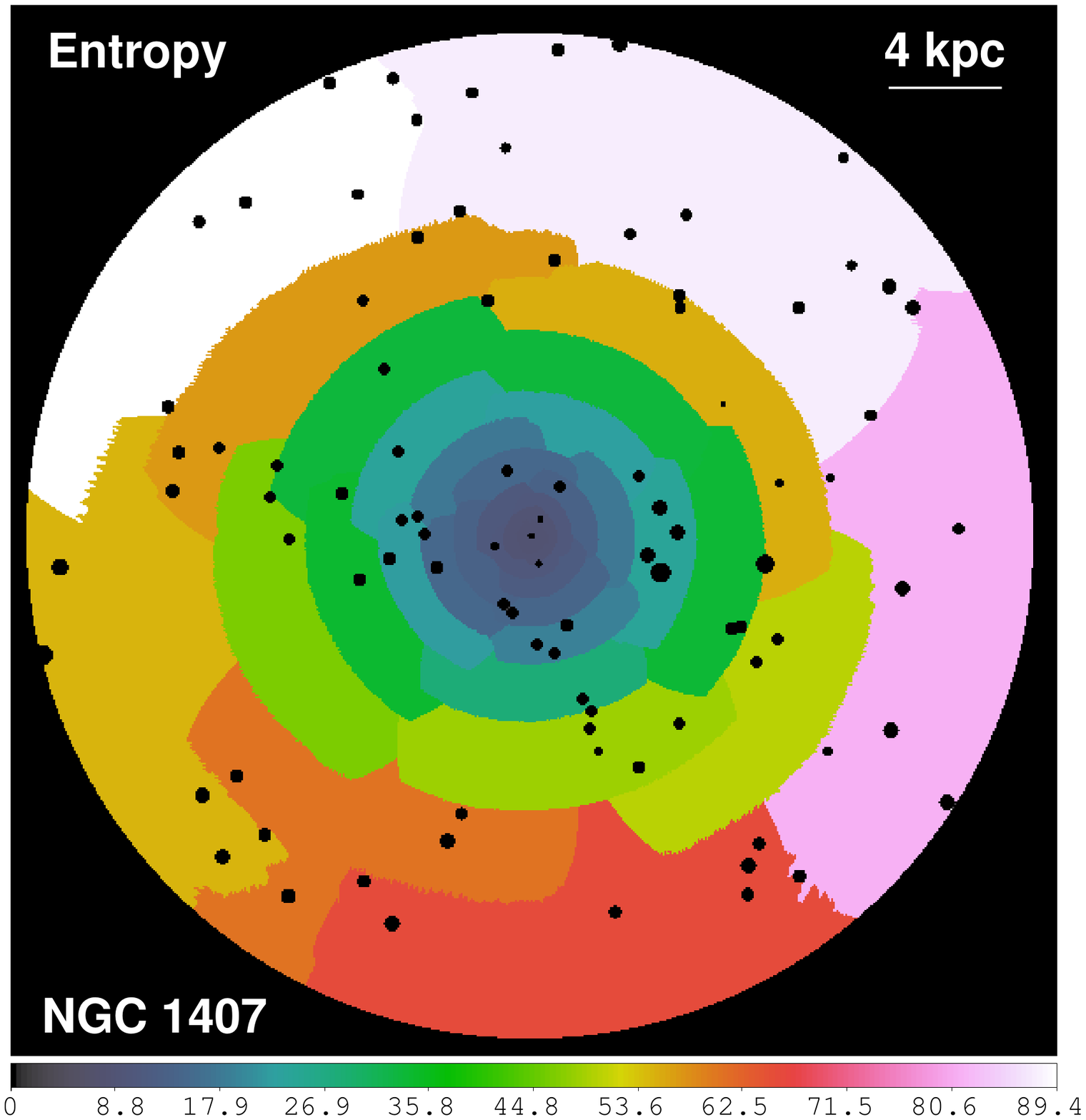}
\end{minipage}
\begin{minipage}{0.32\textwidth}
\includegraphics[height=6cm,clip=t,angle=0.,bb=36 109 577 683]{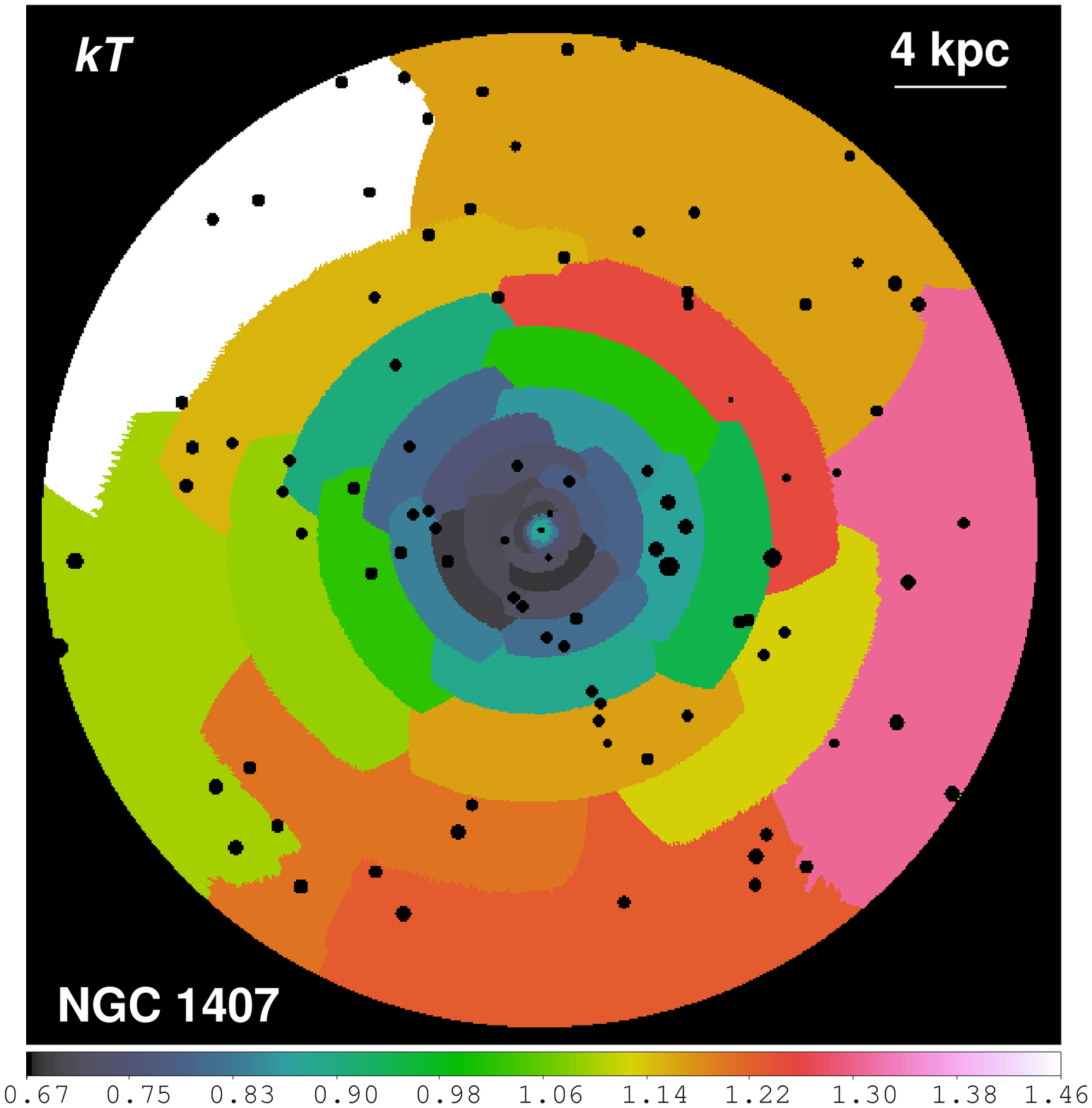}
\end{minipage}
\caption{As Fig. \ref{N4472}, but for NGC~1407. 
}
\label{N1407}
\end{figure*}

Our target list is drawn from the parent sample of \citet{dunn2010}, who identified the optically-brightest ($B_{\rm T}\leq13.5$~mag) and X-ray brightest ($F_{\rm X} > 10^{-12}$~erg\, cm$^{-2}$\, s$^{-1}$ in the 0.1--2.4~keV band) giant elliptical galaxies within a distance $d\leq100$~Mpc and with declination $dec\geq-45$ degrees. Motivated by a goal to resolve the innermost thermodynamic structure in detail, from this list we select only those systems with distance $d<35$~Mpc. 
From the resulting sample of thirteen galaxies, we select those systems that appear the most morphologically relaxed at X-ray wavelengths (i.e. symmetric and undisturbed in their \chandra\ images). The highly symmetric morphologies of these galaxies ensure that their deprojected thermodynamic properties can be determined as robustly as possible. Five galaxies meet these criteria. Their X-ray, near-infrared, and radio images (except for NGC~1407, for which we do not have spatially resolved radio data) are shown in Figs. \ref{N4472}--\ref{N1407} and their properties are summarized in Table~\ref{obs}. 

As seen in the X-ray images (the reduction of the X-ray data is described in Sect.~\ref{reduction}), NGC~4649 and NGC~1407 are highly symmetric. Although NGC~1399, NGC~4472, and NGC~4261 do display slight departures from spherical symmetry, they are still among the most X-ray morphologically relaxed giant elliptical galaxies known. The faintest galaxy in our sample is NGC~4261, with a bolometric X-ray luminosity of ${L_{\rm X}} = 1.63\times 10^{41}$~erg\, s$^{-1}$ and flux of $1.36\times10^{-12}$~erg~s$^{-1}$~cm$^{-2}$. Our relatively high luminosity limit restricts the contamination of the detected X-ray emission by an unresolved population of X-ray binaries (see Sect.~\ref{spectral} for details). The relatively high X-ray fluxes of our sources, combined with their proximity ensure that we can measure the thermodynamic properties down to sub-kpc radii. 

The processed near-infrared K-band images obtained with the Two Micron All-Sky Survey (2MASS) shown in the upper central panels of Figs.~\ref{N4472}--\ref{N1407} were downloaded from the NASA/IPAC Infrared Science Archive\footnote{http://irsa.ipac.caltech.edu/} \citep{jarrett2003}. They show that the stellar light and by implication the star-mass distribution in all of the galaxies is highly symmetric with no indications for disturbance or star-formation. The galaxies do not, in general, show detectable dust lanes or \ion{H}{i} emission and their stellar populations are old and red \citep[][]{serra2010}. NGC~4261 does show a nuclear disk of gas and dust in the innermost $r\sim100$~pc \citep[see][]{jaffe1993,jaffe1994} containing ($5.4\pm1.8$)$\times10^{4}$~$M_{\odot}$ of molecular and atomic hydrogen \citep{ferrarese1996}; however, cool gas or dust has not been detected elsewhere in NGC~4261 \citep{serra2010,combes2007}. All five galaxies are essentially `red and dead'.

All five galaxies have central, active radio jets, but their observed radio powers and morphologies span a wide range, as indicated by the radio luminosities listed in Table~\ref{obs} and by the radio maps in Figs. \ref{N4472}--\ref{N4261} \citep[see also][]{dunn2010}. NGC~4261 is the most powerful radio source, with jets that puncture through the X-ray emitting galactic atmosphere to form giant lobes in the surrounding intra-group medium (IGM). NGC~1399 has $\sim$10~kpc long radio jets with lobes that also appear to be depositing energy away from the centre of the galaxy. The radio lobes of NGC~4472, and of the 10 times less radio luminous NGC~4649, are seen at smaller radii. For NGC~1407, we do not presently have high quality radio data, but its radio luminosity is similar to those of NGC~4472 and NGC~1399 (see Table~\ref{obs}), clearly indicating the presence of jets.

\subsection{\chandra\ data reduction and analysis}
\subsubsection{Data reduction}
\label{reduction}

Our analysis of the {\it Chandra} data follows the data reduction procedures described in \citet{million2010a} and \citet{million2010b}.  
The standard level-1 event lists were reprocessed using the CIAO (version 4.3) software package, including the latest gain maps and calibration products. Bad pixels were removed and standard grade selections applied. The data were cleaned to remove periods of anomalously high background. The net exposure times after cleaning are summarized in Table \ref{obs}. Background images and spectra were extracted from the blank-sky fields available from the Chandra X-ray Center. These were cleaned in an identical way to the source observations, reprojected to the same coordinate system, and normalized by the ratio of the observed to blank-sky count rates in the 9.5--12~keV band. 

Background subtracted images were created in 6 narrow energy bands, spanning 0.5--2.0 keV. These were flat fielded with respect to the median energy for each image and then co-added to create the broad band X-ray images shown in Figs.~ \ref{N4472}--\ref{N1407}. 

Identification of point sources was performed using the CIAO task {\small WAVDETECT}. These were excised from the images in Figs. \ref{N4472}--\ref{N1407} and substituted by local backgrounds. 
Point sources were also excluded from all regions used for spectral analysis. Separate photon-weighted response matrices and effective area files were constructed for each spectral region. 

\subsubsection{Spectral analysis}
\label{spectral}

Our spectral analysis was separated into two parts. In the first part, 2D maps of projected thermodynamic quantities were made.
The individual regions for the 2D spectral mapping were determined using the Contour Binning algorithm \citep{sanders2006b}, which groups neighboring pixels of similar surface brightness until a desired signal-to-noise threshold is met. In order to have small enough regions to resolve substructure, yet still have enough counts to achieve better than 5 per cent accuracy in the temperature measurements, we adopted a signal-to-noise ratio of 18 ($\sim$320~counts per region).
We modeled the spectra extracted from each spatial region with the {\small SPEX}\footnote{www.sron.nl/spex} package \citep[{\small SPEX} is faster, uses an updated version of the MEKAL plasma model with respect to {\small XSPEC}, though provides consistent results]{kaastra1996}. Because the spectral band above 2~keV is dominated by background in low temperature galaxies, the spectral fitting was performed in the 0.6--2.0 keV band. The spectrum for each region was fitted with a model consisting of an absorbed single-phase plasma in collisional ionization equilibrium, with the temperature and spectral normalization (emission measure) as free parameters. The line-of-sight absorption column densities, $N_{\mathrm{H}}$, were fixed to the values determined by the Leiden/Argentine/Bonn radio survey of \ion{H}{i} \citep[][see Table~\ref{obs}]{kalberla2005}. 

After the point source removal, the contribution from an unresolved population of low mass X-ray binaries (LMXB) to the X-ray emission is insignificant. We performed fits, where we included power-law emission components with photon indices $\Gamma=1.56-1.80$, which were shown to describe well the population of low-luminosity LMXB \citep{irwin2003,kim2004}. Such additional power-law emission is not detected significantly in the galaxies and including it in the spectral fitting does not significantly affect the measured thermodynamic properties of the hot ISM (we note that this conclusion does not change if we extend the fitted band to 7 keV). 

The largest systematic uncertainty in the measured densities stems from the metallicity measurements. When fitting the spectra of thermal plasmas with $kT$ in the range $\sim$0.5--1.0~keV, the emission measure and metallicity typically anti-correlate. In regions with complicated temperature structure the metallicity is often biased low and consequently the emission measure is biased high \citep[e. g.][]{buote2000,werner2008}. Moreover, for spectra with relatively low numbers of counts, the statistical uncertainties on the metallicity are high and multi-temperature model fits are unfeasible. Therefore, our 2D maps of thermodynamic properties were produced with the metallicity fixed to 0.5 Solar \citep[throughout the paper, metal abundances are given with respect to the Solar values by][]{grevesse1998}. If the metallicity is underestimated/overestimated by a factor of 2 in the model, the density will be overestimated/underestimated by a factor of $\sim$1.35. 

In the second stage of our analysis, we measured azimuthally-averaged, deprojected radial profiles of thermodynamic quantities. We extracted spectra from concentric annular regions at least 1~arcsec wide, with a signal-to-noise ratio of at least 18 ($\sim$320 counts). The deprojected profiles of thermodynamic properties were obtained with the {\small XSPEC} \citep[version 12.5][]{arnaud1996} spectral fitting package, using the {\small projct} model. The combined set of azimuthally averaged spectra extracted from concentric annuli was modeled in the 0.6--2.0~keV band simultaneously to determine the deprojected electron density ($n_{\rm e}$) and temperature ($kT_{\rm e}$) profiles. The emission from each spherical shell was modeled with a photoelectrically absorbed \citep{balucinska1992} APEC thermal plasma model \citep[][using AtomDB v2.0.1]{smith2001}. Deprojected densities were determined in 14--27 shells, and temperatures in 7--10 regions, depending on the galaxy. For each galaxy we fitted two metallicity values, one for the shells outside the radius of $\sim$1 kpc and another value for the shells in the central regions of the galaxies.

For all spectral fitting, we employ the extended C-statistics available in {\small XSPEC} and {\small SPEX}. All errors are quoted at the 68 per cent confidence level for one interesting parameter  ($\Delta \rm C=1$).

\section{Thermodynamic measurements}
\label{results}

\subsection{2D distribution of thermodynamic properties}

The lower panels of Figs. \ref{N4472}--\ref{N1407} show 2D maps of projected electron pressures ($n_{\rm e}kT_{\rm e}$), entropies ($kT_{\rm e}/n_{\rm e}^{\frac{2}{3}}$), and temperatures ($kT_{\rm e}$). The projected electron densities were determined assuming a fixed line of sight of column depth $l=20$~kpc. 

The 2D distributions of the pressures are approximately spherically symmetric in the cores of all five galaxies, consistent with the X-ray emitting gas being in approximate hydrostatic equilibrium. The most obvious departure from spherical symmetry is seen in NGC~4472 in the areas spatially coincident with the radio lobes at radii $r\sim4$~kpc, where the projected thermal pressure drops by 10--25 per cent.

The spatial distribution of the projected entropy appears spherically symmetric in 4/5 systems, the exception being NGC~1399 where the entropy distribution beyond $r\sim1.5$~kpc is elongated, with lower entropy gas extended along the radio jets. Overall, however, in all cases, the lowest entropy gas resides in the innermost cores of the galaxies, as expected for convective stability. The regions of relatively high entropy at $r>6$~kpc, to the north of NGC~4472 and NGC~1407, and to the east of NGC~1399, indicate that these galaxies are moving through the ambient medium. 

The observed temperature distributions of the galaxies appear less symmetric. In NGC~4472 we see cooler plasma extended along the eastern radio lobe, and in NGC~1399 the low temperature ISM is distributed along the radio jets. 
The lowest projected temperatures of the galactic atmospheres are in the range 0.6--0.75~keV. Such low ISM temperatures are not, however, always seen in the densest, central regions. Three out of five galaxies (NGC~4649, NGC~1407, and NGC~1399) show a significant temperature {\it increase} in their cores. The clearest central temperature increase is seen in NGC~4649, arguably the most morphologically relaxed galaxy in our sample.

\subsection{Azimuthally-averaged, deprojected thermodynamic properties}
\label{deprojected}

The relatively undisturbed X-ray morphologies and high degree of azimuthal symmetry observed in the distributions of projected thermodynamic quantities confirms that the five systems in this study are among the most dynamically relaxed giant elliptical galaxies in the local Universe. All systematic uncertainties associated with deprojection analyses, carried out under the assumption of spherical symmetry and of a single-phase nature at each radius, are therefore minimized for these systems. We have determined the azimuthally-averaged deprojected radial profiles of electron density ($n_{\rm e}$), temperature ($kT_{\rm e}$), and metallicity ($Z$) from the \chandra\ data. Using these quantities we also determine the deprojected radial profiles for the cumulative gas mass, entropy ($K=kT_{\rm e}/n_{\rm e}^{\frac{2}{3}}$), electron pressure ($P_{\rm e}=n_{\rm e}kT_{\rm e}$), and cooling time. We define the cooling time as the gas enthalpy divided by the energy radiated per unit volume of the plasma:
\begin{equation}
t_{\mathrm{cool}}=\frac{\frac{5}{2}(n_{\mathrm{e}}+n_{\mathrm{i}})kT}{n_{\mathrm{e}}n_{\mathrm{i}}\Lambda(T)},
\end{equation}
where the ion number density $n_{\mathrm{i}}=0.92n_{\mathrm{e}}$, and $\Lambda(T)$ is the cooling function for Solar metallicity tabulated by \citet{schure2009}. Cooling functions based on older plasma codes \citep{sutherland1993} predict a 9 per cent lower cooling rate for 0.5~keV plasma. 

Our main result from the deprojection analysis is that beyond the central $r\sim1$~kpc, and out to the radii $r\lesssim6$~kpc, the cumulative gas mass, density, pressure, entropy, and cooling time distributions for the X-ray emitting gas follow remarkably similar radial profiles (see Fig.~\ref{profiles}). The observed cumulative gas mass within $r=8$~kpc is $\sim5.8\times10^8~M_{\odot}$ with a system-to-system dispersion of 10 per cent. The entropy profiles follow a simple, power-law form $K\propto r^{\alpha}$ with an index $\alpha=0.92$-1.07. The weighted system-to-system dispersions in entropy at $r=1$~kpc and at $r=5$~kpc are only 14~per cent and $17$~per cent, respectively. 
The density profiles follow the form $n_{\rm e}\propto r^{-\beta}$ with an index $\beta=1.1$--1.46 and with a system-to-system dispersion of 10 per cent at $r=5$~kpc. The pressure profiles have an approximate power-law form of $P_{\rm e}\propto r^{-\gamma}$ with an index $\gamma=0.93$--1.43 and with a system-to-system dispersion of 13 per cent at $r=5$~kpc. The thermodynamic profiles of NGC~4261 can be well described with power-law relations all the way down to the smallest radii measured. 

In the absence of heating, cooling in the central regions of these galaxies would be strong: within the radius $r\lesssim4$~kpc, the cooling times in all five galaxies are $t_{\rm cool}<1$~Gyr. Within $r\lesssim1$~kpc of the center, the cooling times are $t_{\rm cool}<10^8$~yr, and less than a few $10^7$~yr within $0.5$~kpc. 

All five galaxies show a flattening of their central entropy profiles within $r\lesssim1$~kpc. In contrast to the remarkable similarity in the thermodynamic profiles at larger radii ($1\lesssim r \lesssim6$~kpc), within this innermost $r\lesssim1$~kpc region the entropies of the galaxies also show a significant system-to-system scatter of 40 per cent at $r=0.2$~kpc. The system with the least central flattening of the entropy distribution is NGC~4261, where the entropy drops to $K_{0}=0.88\pm0.02$~keV~cm$^2$ at 0.23~kpc. The largest core entropy of $K_{0}=4.1\pm0.5$~keV~cm$^{2}$ at $r=0.27$~kpc is observed in NGC~1407. 

The temperature profiles of the galaxies exhibit significant system-to-system scatter. Interestingly the temperature profiles for NGC~4649, NGC~1399, and NGC~1407 show clear temperature increase in the central regions, which may be associated with repeated shock activity increasing the entropy of the gas. In contrast, the radial temperature distributions in the centres of NGC~4472 and NGC~4261 are relatively flat. 

The best fit metallicities of the hot ISM within $r\sim1$~kpc are $Z\sim0.5$~Solar, except for NGC~1399 which has a core metallicity of $Z\sim0.85$~Solar. Beyond this innermost region ($1<r<6$~kpc) the metallicities of NGC~4649 and NGC~4261 are $Z\sim0.3$~Solar; NGC~1407 has $Z\sim0.75$~Solar, and the metal abundances of NGC~1399 and NGC~4472 are $Z\sim1$~Solar. The systematic uncertainties on the metallicity measurements of $kT\lesssim1$~keV plasma with CCD type detectors are, however, unfortunately still significant.

To verify that the spectroscopically measured temperatures, and other thermodynamic quantities, determined in azimuthally averaged annular regions are approximately unbiased with respect to the true average values at a given radius, we have compared their results with the average values determined from all bins at the same radii in the 2D maps. We find no significant differences between the values found by fitting spectra extracted from circular annuli and the azimuthally averaged values determined from 2D maps at any radius. 

The obtained deprojected density and  temperature profiles are broadly consistent with the previously published results in the literature \citep[e.g.][]{humphrey2006b,humphrey2009,churazov2010}.

\section{Discussion}
\label{discussion}
 
Beyond the innermost core, at $r\gtrsim1$~kpc and out to at least $r\sim6$~kpc, the X-ray emitting hot gas mass profiles and thermodynamic profiles exhibit remarkable self-similarity. The density, entropy, pressure, and cooling time distributions follow simple, power-law forms. The total gravitating mass at these radii is also similar in all five galaxies, and is dominated by their stellar populations \citep[][]{humphrey2006b,das2010}. The mass fraction of hot X-ray emitting gas in this radial range is small, of the order of $f_{\rm gas}\lesssim2\times10^{-3}$. 
Because the hot ISM represents a minor and essentially dynamically insignificant fraction of the total mass of the galaxies, the observed similarity of the hot gas mass and its thermodynamic profiles is surprising and reflects a deep and long term stability of energy input from the AGN and from the stellar populations. Interestingly, the ratio of the cumulative gas mass within $r<8$~kpc to the total B-band luminosity for the galaxies in our sample spans a relatively small range of $M_{{\rm gas}}/L_{\rm B} = 7.1-11.3 \times10^{-3}~M_\odot/L_\odot$.

\subsection{Universal thermodynamic profiles}

\begin{figure*}
\begin{minipage}{0.49\textwidth}
\includegraphics[height=9.1cm,clip=t,angle=0.,bb=36 119 577 674]{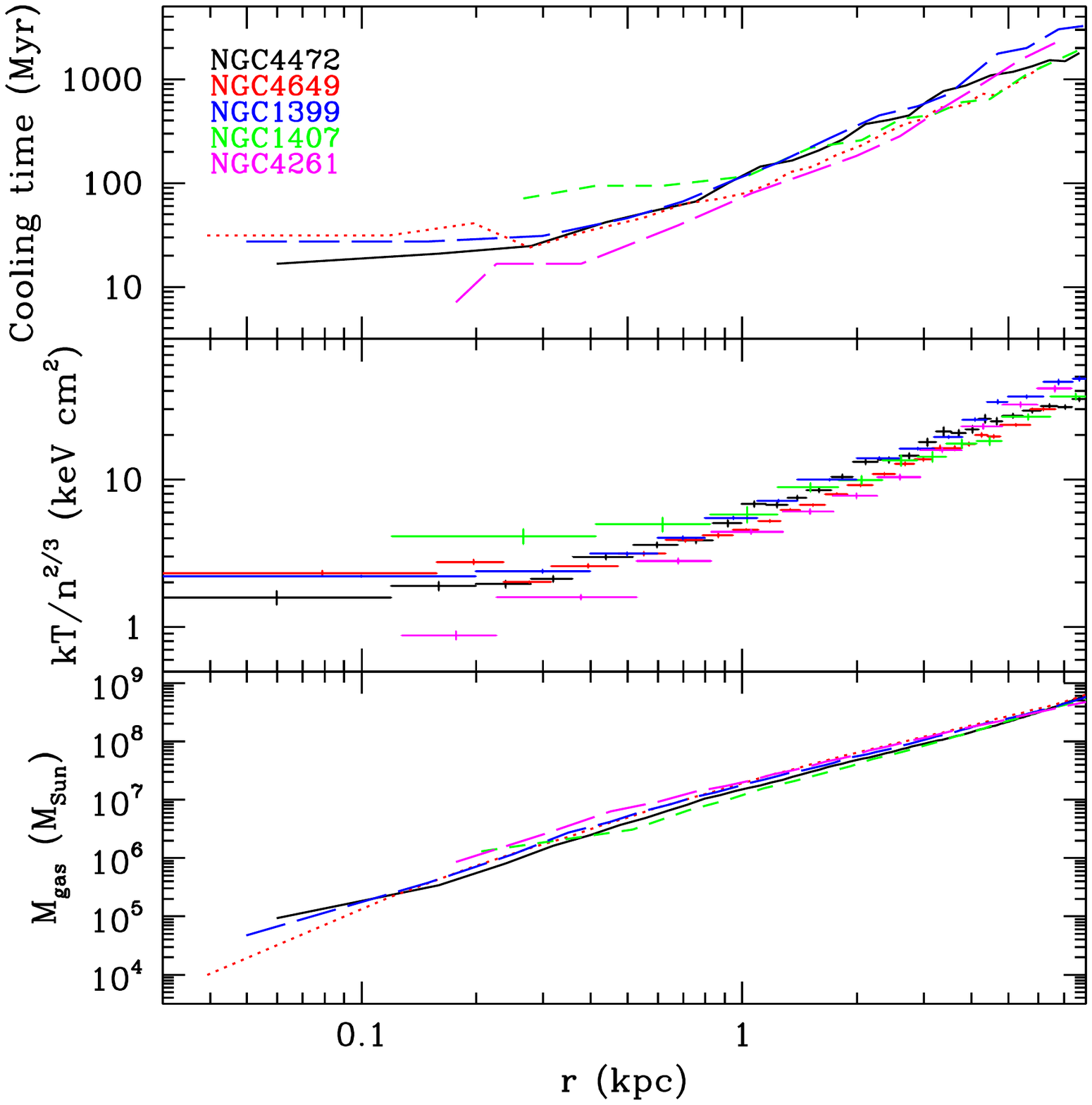}
\end{minipage}
\begin{minipage}{0.49\textwidth}
\includegraphics[height=9.1cm,clip=t,angle=0.,bb=36 119 577 674]{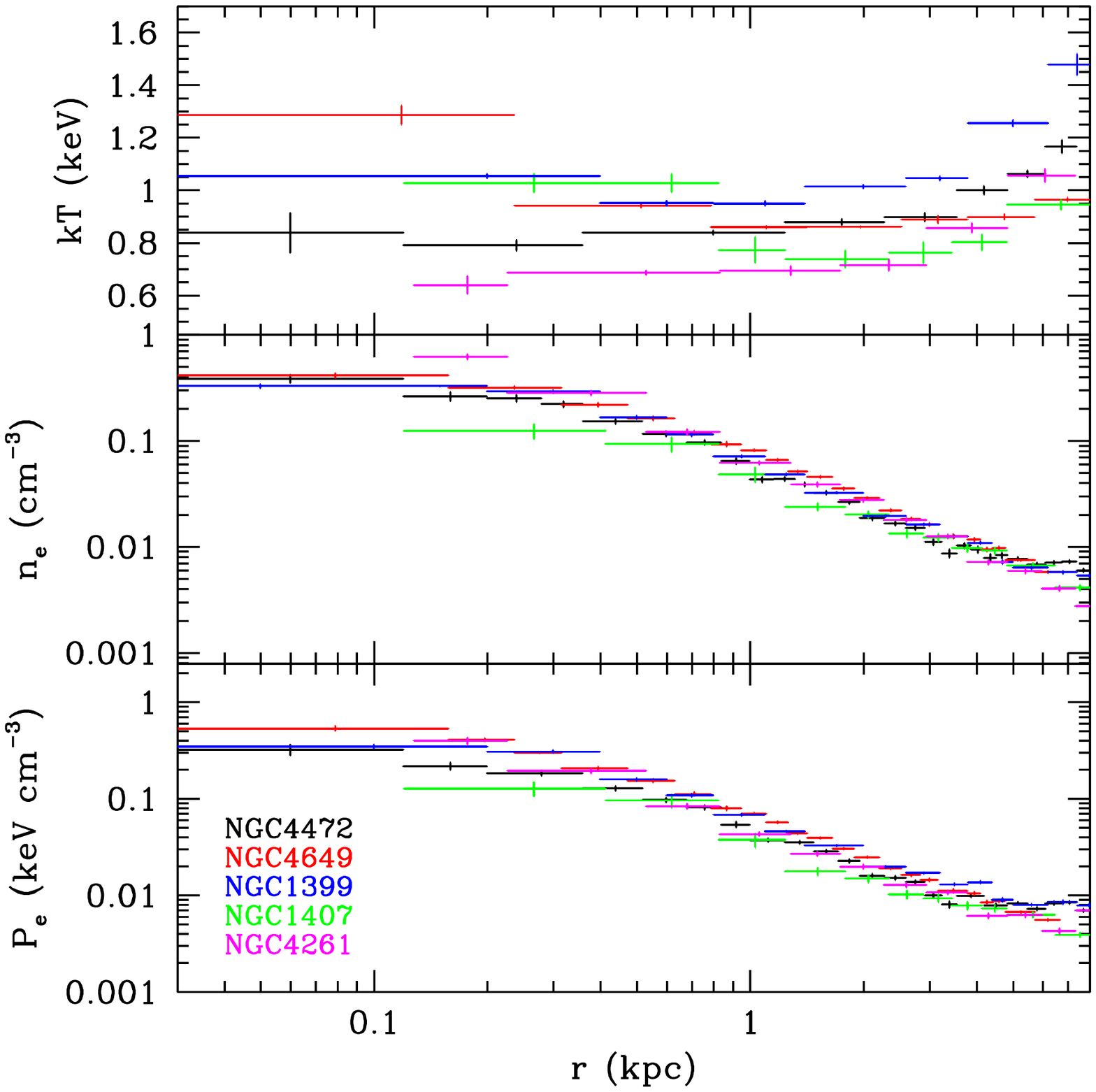}
\end{minipage} \\ \vspace{-0.5cm}
\caption{{\it Left panel:} Radial distributions of cooling times, entropies, and cumulative gas masses for our sample of 5 nearby giant elliptical galaxies. {\it Right panel:} Radial distributions of the deprojected temperatures, electron densities, and electron pressures.  }
\label{profiles}
\end{figure*}

The similarity of the thermodynamic properties of the gaseous atmospheres of our galaxy sample has important implications for the evolution of giant ellipticals. The cooling time of the hot gas in the central regions is short (see left panel of Fig.~\ref{profiles}) and in the absence of heating this gas will cool and form stars \citep[e.g.][]{peterson2006}. The X-ray spectra of these galaxies, however, lack the expected strong \ion{O}{vii} lines associated with gas cooling below $T\sim5\times10^6$~K \citep[see e.g.][]{sanders2011}. Moreover, their stellar populations are old, showing no evidence for recent star-formation, and no significant central atomic or molecular reservoirs of cooled gas are detected \citep{serra2010}. All of this indicates that the gas has been kept from cooling.

Type Ia supernovae are expected to explode in giant elliptical galaxies at a rate of 0.166 per 100 yr per $10^{10}~ L_{\rm B\odot}$  \citep{cappellaro1999}, releasing $\sim10^{51}$~ergs of energy per supernova. In principle, this can provide a heating rate of $1.6\times10^{41}$~erg~s$^{-1}$ within $r\lesssim6$~kpc, which is of the same order of magnitude as the X-ray luminosity within the same region. However, to approximately balance the radiative cooling, the energy released by supernovae would have to couple to the X-ray emitting gas with a 100 per cent efficiency, which seems unlikely. The dominant source of heating is instead likely to be the central AGN.

As indicated by the observed radio morphologies, the jets and lobes driven by the central AGN interact with and mechanically heat the ISM out to a radius of several kpc. A complete, flux-limited study of 18 nearby, X-ray bright, optically bright giant elliptical galaxies by \citet{dunn2010} shows that 17/18 systems exhibit some form of current radio activity associated with the central AGN \citep[see also][]{best2005,dunn2006,dunn2008,sun2009}. Such observations imply that active `radio-mode' AGN feedback is not a rare and sporadic phenomenon, but rather represents the default state for large elliptical galaxies. \citet{allen2006} demonstrate a tight correlation between the Bondi accretion rates of hot gas and the observed jet power in such ellipticals. This argues that AGN feedback is likely controlled by the accretion of hot gas, or multiphase gas entrained within it. The similarity of the observed thermodynamic profiles reported here supports this, and argues that relativistic jets, produced by accretion onto the central SMBH, heat the gas at a similar and constant rate averaged over time scales longer than the cooling time, $t_{\rm cool}\gtrsim10^8$ yr. This jet heating creates an energy balance where heating and cooling are in approximate equilibrium, keeping the hot galactic atmospheres in a near `steady-state'. 
Importantly, the observed `steady-state' also indicates that the time averaged accretion rate onto the SMBH is intimately connected to the conditions of the hot X-ray emitting gas phase on spatial scales of several kpc. 

The jet-powers estimated from the energies and timescales required to inflate the bubbles of relativistic radio-emitting plasma are $P_{\rm jet}\sim8\times10^{42}$~ergs~s$^{-1}$ in NGC~4472 \citep{allen2006}; $P_{\rm jet}\sim1.3\times10^{42}$~ergs~s$^{-1}$ in NGC~4649 and $P_{\rm jet}\sim2.2\times10^{42}$~ergs~s$^{-1}$ in NGC~1399 \citep{shurkin2008}; and $P_{\rm jet}>10^{43}$~ergs~s$^{-1}$ in NGC~4261 \citep{osullivan2011}. 
In NGC~4472, NGC~4649, and NGC~1399 the energy required to keep the hot ISM within $r\lesssim6$~kpc from cooling is only 2.5--10 per cent of the estimated jet-power, and in NGC~4261, which is the system with the most powerful radio source in our sample, this energy is less than 1 per cent of the power output of the AGN. In NGC~4261 the jets penetrate through the hot galactic corona and form lobes which deposit most of their energy into the surrounding intergalactic medium \citep{osullivan2011}. 

Although we do not see evidence for shocks currently propagating through the X-ray atmospheres of the galaxies in our sample, repeated AGN induced shocks most likely do contribute to the heating of the hot ISM. Detailed studies of the giant elliptical galaxy NGC~5813 revealed that the total energy in shocks driven by the expanding bubbles can be 1/2--3 times the total internal energy of the cavities \citep{randall2011}. The total jet power in these systems is therefore most likely even higher than the estimates provided above. Interestingly, despite the fact that the fraction of the estimated `current' jet-power  that is needed in each case to keep the hot ISM from cooling varies from galaxy to galaxy by an order of magnitude, the observed similarity of the thermodynamic properties, including the gas mass and entropy, argues that the energy deposited into the ISM within $r\lesssim6$~kpc must be similar and steady, to keep the hot galactic atmospheres in a `steady-state'.

\subsection{Heating in the innermost $r\lesssim1$~kpc}

As shown in Sect~\ref{deprojected}, the entropy and cooling time profiles show well resolved flattening at radii $r\lesssim1$~kpc. This flattening, combined with the short cooling times of $t_{\rm cool}<10^8$~yr, implies that the gas at these radii has been heated by AGN activity relatively recently. Radiative cooling will reduce the entropy of the central gas, and in order to offset such cooling and flatten the profiles a heating mechanism is required. Heating by repeated AGN induced weak shocks is likely to be the most significant channel for entropy increase of the hot plasma near to the AGN. Shock heating, which increases the entropy by $\Delta S$, will offset the radiative heat loss by $\Delta Q=T\Delta S$ (where $T$ is the temperature of the gas). Examples of pronounced, ongoing AGN induced shock heating have been studied in detail in M~87 \citep{forman2005,forman2007,million2010b} and in NGC~5813 \citep{randall2011}. 

The steepest central entropy distribution is seen in NGC~4261, where the entropy continues to drop all the way in to the centre of the system. The fact that the central entropies at $r<1$~kpc, where $t_{\rm cool}<10^8$~yr, are substantially different from system to system indicates that the accretion rate onto the black hole and the resulting jet-power, heating the ISM, varies significantly on time scales shorter than a few $10^7$~yr.  

Given the remarkable similarity of the profiles at $r>1$~kpc our analysis suggests that the application of hydrostatic mass analyses to the regions of those galaxies from 1--6~kpc are likely to be relatively accurate \citep[e. g.][]{humphrey2006b,churazov2008,churazov2010}. 
\citet{humphrey2008,humphrey2009} extend their hydrostatic mass analyses all the way into the cores of their target galaxies, arguing that in some systems, most notably in NGC~4649, relative hydrostatic equilibrium applies all the way to small radii allowing to infer directly the mass of the central SMBH. 
The increased system-to-system scatter in our observed central gas entropy indicates that caution should be employed when extending such analyses to the inner 1~kpc, where feedback may induce significant thermodynamic variations on timescales as short as a few $10^6$ yr. 

\subsection{The origin of the hot gas}

There are three possible explanations for the origin of the hot gas in giant ellipticals: first, it could be leftover baryonic material from the process of galaxy formation. In this case, the similarity of the gas mass to stellar light ratios in the galaxies would indicate similar star formation and feedback histories. 
Second, some fraction of the X-ray emitting gas could have been accreted from the surrounding large scale environment. If this process is important, then the gas mass to stellar light ratios would depend strongly on the large scale environments of the galaxies and would therefore be expected to exhibit significant system-to-system variation, which is not seen in our sample\footnote{The gas mass fractions around the brightest cluster galaxies of cooling core clusters are significantly higher than those of the galaxies investigated in this paper \citep[e. g.][]{allen2008}. Most of the gas in those systems likely originates from the surrounding intra-cluster medium.}. A third possibility is that stellar mass loss contributes significantly to the X-ray halos \citep[for a review see][]{mathews2003}. In this process, gas from the outer layers of evolved stars is mixed and assimilated into the hot phases of the galaxies \citep{mathews1990}. Hydrodynamic simulations of the gas ejected by red giant stars and planetary nebulae have been performed by \citet{parriott2008} and \citet{bregman2009}. These simulations predict that about 75 per cent of the ejecta produced by red giant stars moving supersonically relative to the ambient medium will be shock heated to approximately the temperature of the hot ISM. 
However, the products of stellar mass loss evidently do not thermalize universally in all environments. The presence of dust and PAHs in the dense cooling cores of some galaxy clusters suggests that in those systems much of the ejected stellar gas remains cool \citep{voitdonahue2011,donahue2011}. The high ambient pressures in the cores of galaxy clusters with massive cooling cores might be promoting more rapid radiative cooling of the products of stellar mass loss. However, none of our galaxies lie at the centers of massive cooling cores.

While we cannot at present discriminate clearly whether original baryonic matter left over from the process of galaxy formation or the products of stellar mass loss dominate the X-ray emitting material within $1<r<8$~ kpc, the remarkable similarity of the thermodynamic and gas mass profiles, and gas mass to stellar light ratios in the systems presented here provides an important, new constraint.

\section{Conclusions}
\label{conclusions}

We performed a detailed spatially resolved study of the thermodynamic properties of the X-ray emitting gas in the inner regions of the five nearest, X-ray and optically brightest, and most morphologically relaxed giant elliptical galaxies at X-ray wavelengths. Our main conclusions may be summarized as follows:

\begin{itemize}

\item Beyond the innermost central region, at $r\gtrsim1$~kpc and out to $r\sim6$~kpc, the radial profiles of density, pressure, cooling time, and entropy of these galaxies follow remarkably similar, simple, power-law like distributions. The entropy profiles follow a form $K\propto r^{\alpha}$ with an index $\alpha=0.92$-1.07, with a system to system dispersion at a given radius of only $\sim$15 per cent. The cumulative hot X-ray emitting gas mass profiles and the gas-mass to stellar-light ratios of all five galaxies are also similar.

\item The observed similarity of the thermodynamic profiles in this radial range argues that, in these systems, relativistic jets heat the gas at a similar rate averaged over time scales longer than the cooling time $t_{\rm cool}\gtrsim10^8$ yr. This jet heating creates an energy balance where heating and cooling are in equilibrium, keeping the hot galactic atmospheres in a `steady-state'.

\item The entropy profiles show well resolved flattening at radii $r\lesssim1$~kpc. In contrast to the remarkable similarity at larger radii ($1\lesssim r\lesssim 6$~kpc), the central entropy value differs from system to system, with a scatter of 40 per cent at 0.2~kpc. The accretion rate onto the black hole and the AGN activity, heating the ISM, must therefore vary significantly on time scales shorter than $t_{\rm cool}=10^7$--$10^8$~yr.  

\item Our results support the picture in which the jets associated with the central AGN are powered by the accretion of hot gas, or the material entrained within it.

\end{itemize}

\section*{Acknowledgments}
We thank Robert Dunn for providing us the radio data for NGC~4472, NGC~4649, and NGC~1399. We thank Payel Das for providing us stellar mass and dark matter profiles for the galaxies. We thank Paul Nulsen and Mark Voit for inspiring discussions. 
Support for this work was provided by the National Aeronautics and Space Administration through Chandra/Einstein Postdoctoral Fellowship Award Number PF8-90056 and PF9-00070 and through the Chandra Award Number GO9-0088X issued by the Chandra X-ray Observatory Center, which is operated by the Smithsonian Astrophysical Observatory for and on behalf of the National Aeronautics and Space Administration under contract NAS8-03060. SWA acknowledges support from the U.S. Department of Energy under contract number DE-AC02-76SF00515 

\bibliographystyle{mnras}
\bibliography{clusters}

\begin{thebibliography}{74}
\expandafter\ifx\csname natexlab\endcsname\relax\def\natexlab#1{#1}\fi

\bibitem[{Allen} et~al.(2006){Allen}, {Dunn}, {Fabian}, {Taylor} \&
  {Reynolds}]{allen2006}
{Allen} S.~W., {Dunn} R.~J.~H., {Fabian} A.~C., {Taylor} G.~B., {Reynolds}
  C.~S., 2006, \mnras, 372, 21

\bibitem[{Allen} et~al.(2008){Allen}, {Rapetti}, {Schmidt}, {Ebeling}, {Morris}
  \& {Fabian}]{allen2008}
{Allen} S.~W., {Rapetti} D.~A., {Schmidt} R.~W., {Ebeling} H., {Morris} R.~G.,
  {Fabian} A.~C., 2008, \mnras, 383, 879

\bibitem[{Arnaud}(1996)]{arnaud1996}
{Arnaud} K.~A., 1996, in { Astronomical Data Analysis Software and Systems
  V\/}, edited by G.~H. {Jacoby}, J.~{Barnes}, vol. 101 of { Astronomical
  Society of the Pacific Conference Series\/}, ~17

\bibitem[{B{\^ i}rzan} et~al.(2004){B{\^ i}rzan}, {Rafferty}, {McNamara},
  {Wise} \& {Nulsen}]{birzan2004}
{B{\^ i}rzan} L., {Rafferty} D.~A., {McNamara} B.~R., {Wise} M.~W., {Nulsen}
  P.~E.~J., 2004, \apj, 607, 800

\bibitem[{Balucinska-Church} \& {McCammon}(1992)]{balucinska1992}
{Balucinska-Church} M., {McCammon} D., 1992, \apj, 400, 699

\bibitem[{Best} et~al.(2005){Best}, {Kauffmann}, {Heckman} et~al.]{best2005}
{Best} P.~N., {Kauffmann} G., {Heckman} T.~M., et~al., 2005, \mnras, 362, 25

\bibitem[{Blakeslee} et~al.(2009){Blakeslee}, {Jord{\'a}n}, {Mei}
  et~al.]{blakeslee2009}
{Blakeslee} J.~P., {Jord{\'a}n} A., {Mei} S., et~al., 2009, \apj, 694, 556

\bibitem[{Bregman} \& {Parriott}(2009)]{bregman2009}
{Bregman} J.~N., {Parriott} J.~R., 2009, \apj, 699, 923

\bibitem[{Buote}(2000)]{buote2000}
{Buote} D.~A., 2000, \mnras, 311, 176

\bibitem[{Cappellaro} et~al.(1999){Cappellaro}, {Evans} \&
  {Turatto}]{cappellaro1999}
{Cappellaro} E., {Evans} R., {Turatto} M., 1999, \aap, 351, 459

\bibitem[{Cavagnolo} et~al.(2009){Cavagnolo}, {Donahue}, {Voit} \&
  {Sun}]{cavagnolo2009}
{Cavagnolo} K.~W., {Donahue} M., {Voit} G.~M., {Sun} M., 2009, \apjs, 182, 12

\bibitem[{Churazov} et~al.(2000){Churazov}, {Forman}, {Jones} \&
  {B{\"o}hringer}]{churazov2000}
{Churazov} E., {Forman} W., {Jones} C., {B{\"o}hringer} H., 2000, \aap, 356,
  788

\bibitem[{Churazov} et~al.(2008){Churazov}, {Forman}, {Vikhlinin}, {Tremaine},
  {Gerhard} \& {Jones}]{churazov2008}
{Churazov} E., {Forman} W., {Vikhlinin} A., {Tremaine} S., {Gerhard} O.,
  {Jones} C., 2008, \mnras, 388, 1062

\bibitem[{Churazov} et~al.(2010){Churazov}, {Tremaine}, {Forman}
  et~al.]{churazov2010}
{Churazov} E., {Tremaine} S., {Forman} W., et~al., 2010, \mnras, 404, 1165

\bibitem[{Combes} et~al.(2007){Combes}, {Young} \& {Bureau}]{combes2007}
{Combes} F., {Young} L.~M., {Bureau} M., 2007, \mnras, 377, 1795

\bibitem[{Condon} et~al.(2002){Condon}, {Cotton} \& {Broderick}]{condon2002}
{Condon} J.~J., {Cotton} W.~D., {Broderick} J.~J., 2002, \aj, 124, 675

\bibitem[{Condon} et~al.(1998){Condon}, {Cotton}, {Greisen} et~al.]{condon1998}
{Condon} J.~J., {Cotton} W.~D., {Greisen} E.~W., et~al., 1998, \aj, 115, 1693

\bibitem[{Croton} et~al.(2006){Croton}, {Springel}, {White} et~al.]{croton2006}
{Croton} D.~J., {Springel} V., {White} S.~D.~M., et~al., 2006, \mnras, 365, 11

\bibitem[{Das} et~al.(2010){Das}, {Gerhard}, {Churazov} \&
  {Zhuravleva}]{das2010}
{Das} P., {Gerhard} O., {Churazov} E., {Zhuravleva} I., 2010, \mnras, 409, 1362

\bibitem[{De Lucia} \& {Blaizot}(2007)]{delucia2007}
{De Lucia} G., {Blaizot} J., 2007, \mnras, 375, 2

\bibitem[{Donahue} et~al.(2011){Donahue}, {de Messi{\`e}res}, {O'Connell}
  et~al.]{donahue2011}
{Donahue} M., {de Messi{\`e}res} G.~E., {O'Connell} R.~W., et~al., 2011, \apj,
  732, 40

\bibitem[{Dunn} et~al.(2010){Dunn}, {Allen}, {Taylor} et~al.]{dunn2010}
{Dunn} R.~J.~H., {Allen} S.~W., {Taylor} G.~B., et~al., 2010, \mnras

\bibitem[{Dunn} \& {Fabian}(2006)]{dunn2006}
{Dunn} R.~J.~H., {Fabian} A.~C., 2006, \mnras, 373, 959

\bibitem[{Dunn} \& {Fabian}(2008)]{dunn2008}
{Dunn} R.~J.~H., {Fabian} A.~C., 2008, \mnras, 385, 757

\bibitem[{Dunn} et~al.(2005){Dunn}, {Fabian} \& {Taylor}]{dunn2005}
{Dunn} R.~J.~H., {Fabian} A.~C., {Taylor} G.~B., 2005, \mnras, 364, 1343

\bibitem[{Fabian} et~al.(2003){Fabian}, {Sanders}, {Allen} et~al.]{fabian2003}
{Fabian} A.~C., {Sanders} J.~S., {Allen} S.~W., et~al., 2003, \mnras, 344, L43

\bibitem[{Fabian} et~al.(2006){Fabian}, {Sanders}, {Taylor} et~al.]{fabian2006}
{Fabian} A.~C., {Sanders} J.~S., {Taylor} G.~B., et~al., 2006, \mnras, 366, 417

\bibitem[{Ferrarese} et~al.(1996){Ferrarese}, {Ford} \& {Jaffe}]{ferrarese1996}
{Ferrarese} L., {Ford} H.~C., {Jaffe} W., 1996, \apj, 470, 444

\bibitem[{Forman} et~al.(2007){Forman}, {Jones}, {Churazov} et~al.]{forman2007}
{Forman} W., {Jones} C., {Churazov} E., et~al., 2007, \apj, 665, 1057

\bibitem[{Forman} et~al.(2005){Forman}, {Nulsen}, {Heinz} et~al.]{forman2005}
{Forman} W., {Nulsen} P., {Heinz} S., et~al., 2005, \apj, 635, 894

\bibitem[{Gitti} et~al.(2012){Gitti}, {Brighenti} \& {McNamara}]{gitti2012}
{Gitti} M., {Brighenti} F., {McNamara} B.~R., 2012, Advances in Astronomy, 2012

\bibitem[{Grevesse} \& {Sauval}(1998)]{grevesse1998}
{Grevesse} N., {Sauval} A.~J., 1998, Space Science Reviews, 85, 161

\bibitem[{Humphrey} et~al.(2008){Humphrey}, {Buote}, {Brighenti}, {Gebhardt} \&
  {Mathews}]{humphrey2008}
{Humphrey} P.~J., {Buote} D.~A., {Brighenti} F., {Gebhardt} K., {Mathews}
  W.~G., 2008, \apj, 683, 161

\bibitem[{Humphrey} et~al.(2009){Humphrey}, {Buote}, {Brighenti}, {Gebhardt} \&
  {Mathews}]{humphrey2009}
{Humphrey} P.~J., {Buote} D.~A., {Brighenti} F., {Gebhardt} K., {Mathews}
  W.~G., 2009, \apj, 703, 1257

\bibitem[{Humphrey} et~al.(2006){Humphrey}, {Buote}, {Gastaldello}
  et~al.]{humphrey2006b}
{Humphrey} P.~J., {Buote} D.~A., {Gastaldello} F., et~al., 2006, \apj, 646, 899

\bibitem[{Irwin} et~al.(2003){Irwin}, {Athey} \& {Bregman}]{irwin2003}
{Irwin} J.~A., {Athey} A.~E., {Bregman} J.~N., 2003, \apj, 587, 356

\bibitem[{Jaffe} et~al.(1993){Jaffe}, {Ford}, {Ferrarese}, {van den Bosch} \&
  {O'Connell}]{jaffe1993}
{Jaffe} W., {Ford} H.~C., {Ferrarese} L., {van den Bosch} F., {O'Connell}
  R.~W., 1993, \nat, 364, 213

\bibitem[{Jaffe} \& {McNamara}(1994)]{jaffe1994}
{Jaffe} W., {McNamara} B.~R., 1994, \apj, 434, 110

\bibitem[{Jarrett} et~al.(2003){Jarrett}, {Chester}, {Cutri}, {Schneider} \&
  {Huchra}]{jarrett2003}
{Jarrett} T.~H., {Chester} T., {Cutri} R., {Schneider} S.~E., {Huchra} J.~P.,
  2003, \aj, 125, 525

\bibitem[{Kaastra} et~al.(1996){Kaastra}, {Mewe} \&
  {Nieuwenhuijzen}]{kaastra1996}
{Kaastra} J.~S., {Mewe} R., {Nieuwenhuijzen} H., 1996, in { UV and X-ray
  Spectroscopy of Astrophysical and Laboratory Plasmas p.411, K. Yamashita and
  T. Watanabe. Tokyo : Universal Academy Press\/}

\bibitem[{Kalberla} et~al.(2005){Kalberla}, {Burton}, {Hartmann}
  et~al.]{kalberla2005}
{Kalberla} P.~M.~W., {Burton} W.~B., {Hartmann} D., et~al., 2005, \aap, 440,
  775

\bibitem[{Kim} \& {Fabbiano}(2004)]{kim2004}
{Kim} D.-W., {Fabbiano} G., 2004, \apj, 611, 846

\bibitem[{Mathews}(1990)]{mathews1990}
{Mathews} W.~G., 1990, \apj, 354, 468

\bibitem[{Mathews} \& {Brighenti}(2003)]{mathews2003}
{Mathews} W.~G., {Brighenti} F., 2003, ARA\&A, 41, 191

\bibitem[{McNamara} \& {Nulsen}(2007)]{mcnamara2007}
{McNamara} B.~R., {Nulsen} P.~E.~J., 2007, ARA\&A, 45, 117

\bibitem[{Million} et~al.(2010{\natexlab{a}}){Million}, {Allen}, {Werner} \&
  {Taylor}]{million2010a}
{Million} E.~T., {Allen} S.~W., {Werner} N., {Taylor} G.~B.,
  2010{\natexlab{a}}, \mnras, 405, 1624

\bibitem[{Million} et~al.(2010{\natexlab{b}}){Million}, {Werner}, {Simionescu}
  et~al.]{million2010b}
{Million} E.~T., {Werner} N., {Simionescu} A., et~al., 2010{\natexlab{b}},
  \mnras, 407, 2046

\bibitem[{Nulsen} et~al.(2005){Nulsen}, {McNamara}, {Wise} \&
  {David}]{nulsen2005}
{Nulsen} P.~E.~J., {McNamara} B.~R., {Wise} M.~W., {David} L.~P., 2005, \apj,
  628, 629

\bibitem[{O'Sullivan} et~al.(2001){O'Sullivan}, {Forbes} \&
  {Ponman}]{osullivan2001}
{O'Sullivan} E., {Forbes} D.~A., {Ponman} T.~J., 2001, \mnras, 328, 461

\bibitem[{O'Sullivan} et~al.(2011){O'Sullivan}, {Worrall}, {Birkinshaw}
  et~al.]{osullivan2011}
{O'Sullivan} E., {Worrall} D.~M., {Birkinshaw} M., et~al., 2011, \mnras,
  1180--+

\bibitem[{Parriott} \& {Bregman}(2008)]{parriott2008}
{Parriott} J.~R., {Bregman} J.~N., 2008, \apj, 681, 1215

\bibitem[{Peterson} \& {Fabian}(2006)]{peterson2006}
{Peterson} J.~R., {Fabian} A.~C., 2006, Phys.~Rep., 427, 1

\bibitem[{Pratt} et~al.(2010){Pratt}, {Arnaud}, {Piffaretti} et~al.]{pratt2010}
{Pratt} G.~W., {Arnaud} M., {Piffaretti} R., et~al., 2010, \aap, 511, A85+

\bibitem[{Pratt} et~al.(2006){Pratt}, {Arnaud} \& {Pointecouteau}]{pratt2006a}
{Pratt} G.~W., {Arnaud} M., {Pointecouteau} E., 2006, \aap, 446, 429

\bibitem[{Rafferty} et~al.(2006){Rafferty}, {McNamara}, {Nulsen} \&
  {Wise}]{rafferty2006}
{Rafferty} D.~A., {McNamara} B.~R., {Nulsen} P.~E.~J., {Wise} M.~W., 2006,
  \apj, 652, 216

\bibitem[{Randall} et~al.(2011){Randall}, {Forman}, {Giacintucci}
  et~al.]{randall2011}
{Randall} S.~W., {Forman} W.~R., {Giacintucci} S., et~al., 2011, \apj, 726, 86

\bibitem[{Sanders}(2006)]{sanders2006b}
{Sanders} J.~S., 2006, \mnras, 371, 829

\bibitem[{Sanders} \& {Fabian}(2011)]{sanders2011}
{Sanders} J.~S., {Fabian} A.~C., 2011, \mnras, 412, L35

\bibitem[{Schure} et~al.(2009){Schure}, {Kosenko}, {Kaastra}, {Keppens} \&
  {Vink}]{schure2009}
{Schure} K.~M., {Kosenko} D., {Kaastra} J.~S., {Keppens} R., {Vink} J., 2009,
  \aap, 508, 751

\bibitem[{Serra} \& {Oosterloo}(2010)]{serra2010}
{Serra} P., {Oosterloo} T.~A., 2010, \mnras, 401, L29

\bibitem[{Shurkin} et~al.(2008){Shurkin}, {Dunn}, {Gentile}, {Taylor} \&
  {Allen}]{shurkin2008}
{Shurkin} K., {Dunn} R.~J.~H., {Gentile} G., {Taylor} G.~B., {Allen} S.~W.,
  2008, \mnras, 383, 923

\bibitem[{Sijacki} et~al.(2007){Sijacki}, {Springel}, {Di Matteo} \&
  {Hernquist}]{sijacki2007}
{Sijacki} D., {Springel} V., {Di Matteo} T., {Hernquist} L., 2007, \mnras, 380,
  877

\bibitem[{Simionescu} et~al.(2009{\natexlab{a}}){Simionescu}, {Roediger},
  {Nulsen} et~al.]{simionescu2009a}
{Simionescu} A., {Roediger} E., {Nulsen} P.~E.~J., et~al., 2009{\natexlab{a}},
  \aap, 495, 721

\bibitem[{Simionescu} et~al.(2009{\natexlab{b}}){Simionescu}, {Werner},
  {B{\"o}hringer} et~al.]{simionescu2009b}
{Simionescu} A., {Werner} N., {B{\"o}hringer} H., et~al., 2009{\natexlab{b}},
  \aap, 493, 409

\bibitem[{Simionescu} et~al.(2008){Simionescu}, {Werner}, {Finoguenov},
  {B{\"o}hringer} \& {Br{\"u}ggen}]{simionescu2008a}
{Simionescu} A., {Werner} N., {Finoguenov} A., {B{\"o}hringer} H.,
  {Br{\"u}ggen} M., 2008, \aap, 482, 97

\bibitem[{Smith} et~al.(2001){Smith}, {Brickhouse}, {Liedahl} \&
  {Raymond}]{smith2001}
{Smith} R.~K., {Brickhouse} N.~S., {Liedahl} D.~A., {Raymond} J.~C., 2001,
  \apjl, 556, L91

\bibitem[{Sun}(2009)]{sun2009}
{Sun} M., 2009, \apj, 704, 1586

\bibitem[{Sun} et~al.(2009){Sun}, {Voit}, {Donahue}, {Jones}, {Forman} \&
  {Vikhlinin}]{sun2009b}
{Sun} M., {Voit} G.~M., {Donahue} M., {Jones} C., {Forman} W., {Vikhlinin} A.,
  2009, \apj, 693, 1142

\bibitem[{Sutherland} \& {Dopita}(1993)]{sutherland1993}
{Sutherland} R.~S., {Dopita} M.~A., 1993, \apjs, 88, 253

\bibitem[{Tonry} et~al.(2001){Tonry}, {Dressler}, {Blakeslee}
  et~al.]{tonry2001}
{Tonry} J.~L., {Dressler} A., {Blakeslee} J.~P., et~al., 2001, \apj, 546, 681

\bibitem[{Voit} \& {Donahue}(2011)]{voitdonahue2011}
{Voit} G.~M., {Donahue} M., 2011, \apjl, 738, L24

\bibitem[{Werner} et~al.(2008){Werner}, {Durret}, {Ohashi}, {Schindler} \&
  {Wiersma}]{werner2008}
{Werner} N., {Durret} F., {Ohashi} T., {Schindler} S., {Wiersma} R.~P.~C.,
  2008, \ssr, 134, 337

\bibitem[{Werner} et~al.(2010){Werner}, {Simionescu}, {Million}
  et~al.]{werner2010}
{Werner} N., {Simionescu} A., {Million} E.~T., et~al., 2010, \mnras, 407, 2063

\bibitem[{Werner} et~al.(2011){Werner}, {Sun}, {Bagchi} et~al.]{werner2011}
{Werner} N., {Sun} M., {Bagchi} J., et~al., 2011, \mnras, 415, 3369

\end{thebibliography}

\end{document}